\documentclass[preprint2]{aastex6}

\usepackage[varg]{txfonts}
\usepackage{natbib}
\usepackage{morefloats}
\usepackage[utf8]{inputenc}
\usepackage{ textcomp }
\usepackage{ gensymb }  
\usepackage{hyperref}   
\usepackage{float}
\usepackage{multirow}

\begin{document}

	\title{Long-Period Intensity Pulsations in Coronal Loops Explained by Thermal Non-Equilibrium Cycles}

	\author{C.~Froment\altaffilmark{1}}
	
	\affiliation{Institut d'Astrophysique Spatiale, CNRS, Univ. Paris-Sud, Universit\'e Paris-Saclay, B\^at. 121, F-91405 Orsay cedex, France} 
	\altaffiltext{1}{Present address: Institute of Theoretical Astrophysics, University of Oslo, P.O Box 1029, Blindern, NO-0315, Oslo, Norway}
	\email{clara.froment@astro.uio.no} 
	
	\author{F.~Auch\`ere}
	\affiliation{Institut d'Astrophysique Spatiale, CNRS, Univ. Paris-Sud, Universit\'e Paris-Saclay, B\^at. 121, F-91405 Orsay cedex, France} 
	
	\author{G.~Aulanier}
	\affiliation{LESIA, Observatoire de Paris, PSL Research University, CNRS, Sorbonne Universit\'es, UPMC Univ. Paris 06, Univ. Paris Diderot, Sorbonne Paris Cit\'e, 5 place Jules Janssen, F-92195 Meudon, France}
	
	\author{Z.~Miki\'c}
	\affiliation{Predictive Science, Inc., San Diego, CA 92121, USA}
	
	\author{K.~Bocchialini}
	\affiliation{Institut d'Astrophysique Spatiale, CNRS, Univ. Paris-Sud, Universit\'e Paris-Saclay, B\^at. 121, F-91405 Orsay cedex, France} 
	
	\author{E.~Buchlin}
	\affiliation{Institut d'Astrophysique Spatiale, CNRS, Univ. Paris-Sud, Universit\'e Paris-Saclay, B\^at. 121, F-91405 Orsay cedex, France} 
	
	\and 
	\author{J.~Solomon}
	\affiliation{Institut d'Astrophysique Spatiale, CNRS, Univ. Paris-Sud, Universit\'e Paris-Saclay, B\^at. 121, F-91405 Orsay cedex, France}

	\accepted{December, 21 2016}

\begin{abstract}

In solar coronal loops, thermal non-equilibrium (TNE) is a phenomenon that can occur when the heating is both highly-stratified and quasi-constant. 
Unambiguous observational identification of TNE would thus permit to strongly constrain heating scenarios. Up to now, while TNE is the standard interpretation of coronal rain, the long-term periodic evolution predicted by simulations has never been observed yet. However, the detection of long-period intensity pulsations (periods of several hours) has been recently reported with SoHO/EIT, and this phenomenon appears to be very common in loops. 
Moreover, the three intensity-pulsation events that we recently studied with SDO/AIA show strong evidence for TNE in warm loops. 
In the present paper, a realistic loop geometry from LFFF extrapolations is used as input to 1D hydrodynamic simulations. 
Our simulations show that for the present loop geometry, the heating has to be asymmetrical to produce TNE. We analyse in detail one particular simulation that reproduces the average thermal behavior of one of the pulsating loop bundle observed with AIA. 
We compare the properties of this simulation with the properties deduced from the observations.
The magnetic topology of the LFFF extrapolations points to the presence of sites of preferred reconnection at one footpoint, supporting the presence of asymmetric heating.
In addition, we can reproduce the temporal large-scale intensity properties of the pulsating loops.
This simulation further strengthens the interpretation of the observed pulsations as signatures of TNE. This thus gives important information on the heating localization and time scale for these loops.

\end{abstract}
	

\section{Introduction}   
 
Numerical simulations show that with a highly-stratified (i.e. mainly concentrated near the footpoints) and quasi-constant heating, a phenomenon called thermal non-equilibrium (TNE) can govern the dynamics of coronal loops. This particular localisation and time scale of the heating results in a particular response of the plasma: evaporation and condensation cycles with periodic evolution of the temperature and the density \citep[e.g.,][]{kuin1982}. The plasma heated near the footpoints 
evaporates and starts to fill the loop. The temperature increases, followed by the density with a time delay. 
As the loop becomes denser, the heating per mass unit decreases and thus the temperature decreases. The temperature drop increases the radiation losses \citep{rosner1978}. The radiation losses start to overcome the heating locally in the corona as the temperature continues to decline, producing thermal runaway.
Eventually, it produces plasma condensations that fall toward the loop legs because of gravity, before the cycle repeats.
The heating stratification produces an unstable loop where no thermal equilibrium is possible \citep[e.g.,][]{antiochos1999,patsourakos_model_2004,klimchuk2010}. The term TNE was introduced by \citet{karpen_are_2001} to describe both the unstable and cyclical aspect of the system. While the thermal runaway occuring \textit{locally} in the corona and on a short time scale (compared to the loop evolution time scale) has all the characteristics of a thermal instability \citep{parker_instability_1953,field_thermal_1965}, we use the term TNE in the present paper since we focus on the  \textit{global} behavior of the system. In the remainder of the paper, we thus use the term TNE since the cyclical aspect of the phenomenon is central to our argument.

\medskip

In simulations, a state of TNE can be produced with different time scales of the heating, as long as the time delay between two heating events is short compared to the typical cooling time. TNE can be obtained either with impulsive heating with a sufficiently high repetition rate \citep[e.g.,][]{mendoza-briceno2005, karpen2008, susino2010,antolin2010}, or with a strictly steady heating \citep[e.g.,][]{muller2003, muller2004, mok2008, lionello2013, mikic2013, mok2016}. 

The role of TNE in the formation and evolution of prominences \citep{antiochos1991,antiochos1999,antiochos2000, karpen2006, xia_formation_2011,xia_simulating_2014} and coronal rain \citep{muller2003,muller2004,antolin2010, fang2013} is both predicted by simulations and confirmed by observations \citep{schrijver2001, degroof2004, muller2005, antolin2012, vashalomidze2015, antolin2015}. In contrast, the relevance of TNE to the description of warm loops (temperatures about 1~MK) has been questioned \citep{klimchuk2010}. However, recent simulations \citep{lionello2013, mikic2013, winebarger2014, lionello2016, mok2016} show that it should not be ruled out because discrepancies with observations may arise from an oversimplification of the loop geometry used in models, especially in 1D simulations.

\medskip

\citet{auchere2014} reported on the detection of long-period intensity pulsations (periods of several hours) with the Extreme Ultraviolet Imaging Telescope \citep[EIT;][]{eit1995} on board the Solar and Heliospheric Observatory \citep[SoHO;][]{soho1995}. This phenomenon appears to be very common in active regions and in particular in loops. The authors estimate that about half the active regions of the year 2000 underwent this kind of pulsation.

These pulsations are new observational signatures of heating processes in loops.
Three of these intensity pulsation events have been analyzed in detail in \citet{froment2015} using simultaneously the six coronal passbands of  the Atmospheric Imaging Assembly \citep[AIA;][]{boerner2012sdo,lemen2012sdo} on board the \textit{Solar Dynamics Observatory} \citep[SDO;][] {pesnell2012sdo}. We used both the Differential Emission Measure (DEM) analysis diagnostics developed by \citet{guennou2012_1, guennou2012_2, guennou2013} and the time lag analysis of \citet{viall&klimchuk2012} to conclude that these loops show evidence for TNE. This conclusion is reinforced by the pulse-train nature of the observed signals \citep{auchere2016b}. Moreover, \citet{auchere2016a} have recently confirmed the statistical significance of the detections.

The time lag analysis presented in \citet[][see e.g. Figure~9]{froment2015} shows that there is no time delay in the loop intensity between the 171~\AA~and 131~\AA~AIA channels. That implies that the plasma does not cool in average below the temperature of the peak response of the 171 channel, i.e. about 0.8~MK \citep[see zero time lag discussions in][]{bradshaw_patterns_2016, viall_signatures_2016}. This specific plasma response seems to match the evaporation and incomplete condensation cycles seen in the simulations of \citet{mikic2013}. 
The authors present several loops modeled with a 1D hydrodynamic code, using different loop symmetries, loop cross-sections, and heating intensity and geometry profiles. Some of their simulations lead to a TNE state of \textquotedblleft incomplete\textquotedblright~condensations. This term is opposed to \textquotedblleft complete\textquotedblright~condensations which are characteristic of catastrophic cooling events where the temperature, locally in the corona, drops down to chromospheric temperatures. The authors investigate the effect of loop geometry (particularly loop cross-sectional area) on the characteristics of loops and conclude on the importance of using realistic loop geometries in numerical models. In addition, the geometry and the strength of the heating seem to play an important role in producing the different regimes of evaporation and condensation cycles.

\medskip

We focus our present analysis on the main event (the one with the largest excess power in Fourier spectra) studied in \citet{froment2015}. We use a 1D hydrodynamic simulation code in order to study if the scenario of cycles of evaporation and incomplete condensations are consistent with our observations. In Section~\ref{sec:simu}, we present the results of one loop simulation that best reproduces the periods and  the average thermal behavior derived from the DEM and the time lag analysis presented in \citet{froment2015}. This simulation uses a non ad hoc loop geometry from linear force-free field (LFFF) extrapolations. We then test in Section~\ref{sec:obs} if the characteristics of both the heating function chosen and the synthetic intensities from the simulation match the observations of the active region.
Finally, we discuss our results and their implications for coronal heating in Section~\ref{sec:discussion} and Section~\ref{sec:summary}.

\section{1D hydrodynamic simulation}\label{sec:simu}

In \citet{froment2015}, the event detected in NOAA~AR~11499 is referred to as event~1 and is followed from June 03, 2012 18:00 UT to June 10, 2012 04:29 UT. In this active region, we detected intensity pulsations with a period of 9.0~hr in large coronal loops. The aim of this section is to reproduce the main characteristics (long term variations and average behavior of the loop bundle observed with AIA) of the plasma response of this pulsating event, as detailed in \citet{froment2015}. We present here results of a 1D hydrodynamic simulation using a realistic loop geometry from a LFFF extrapolation.

	\subsection{Magnetic field extrapolation}\label{sec:mag_extrapol}
	
	
	\begin{figure*}
		\centering
                 \includegraphics[width=\linewidth]{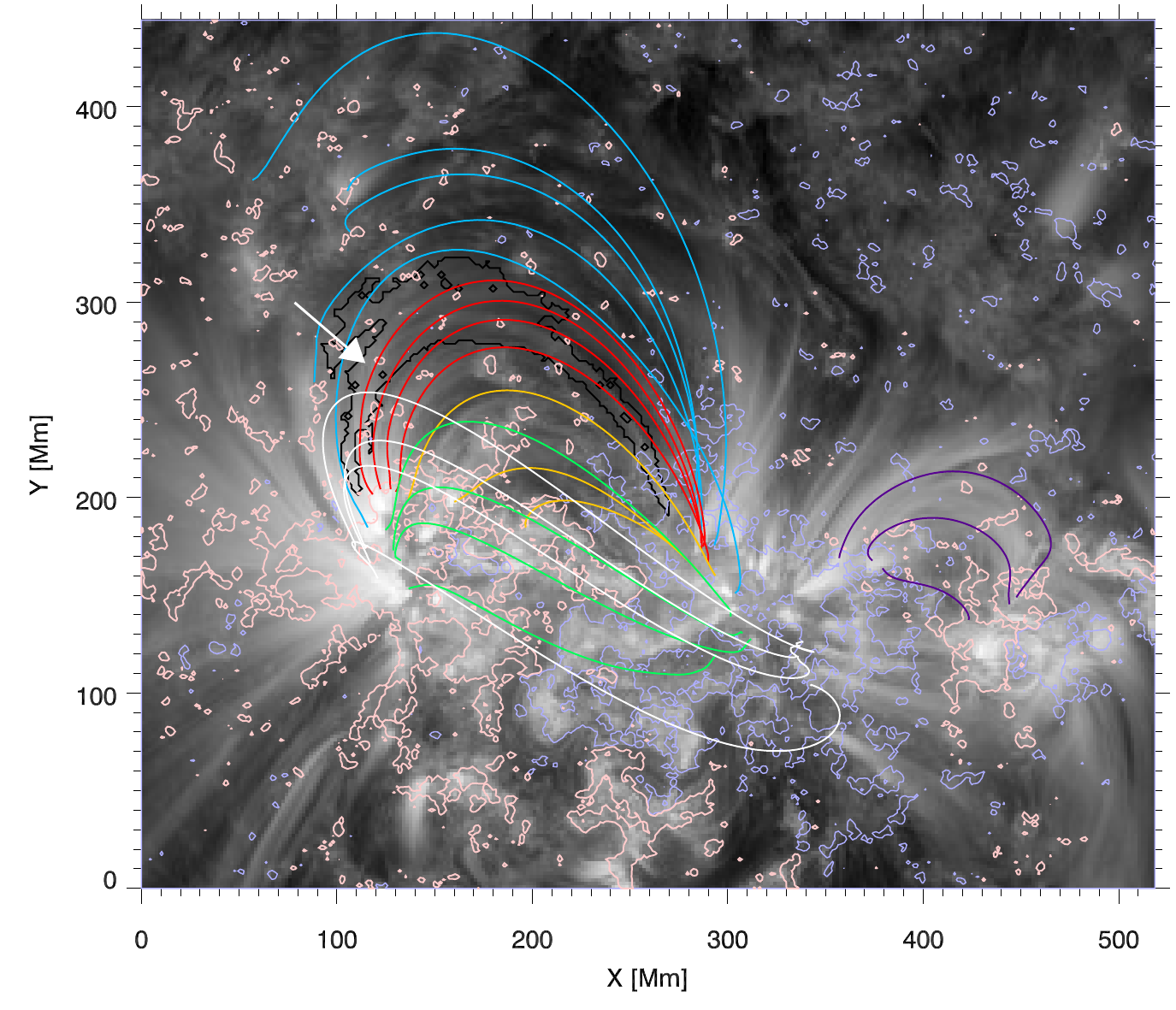}
                 \caption{Some field lines extrapolated for NOAA AR 11499 (and related active regions) with a LFFF model. We superimposed contours of magnetic field ($B_z$ at $z = 0$) from the HMI magnetogram (see Figure~\ref{fig:spe_loopfpt}), in light red for positive values and in light blue for negative ones. The contours represented are at $\pm$30 G. The AIA 171~\AA~image and the magnetogram are both taken on June 06, 2012 at 23:12 UT. They cover a region of 519 Mm along the X-axis and 444 Mm along the Y-axis. The field-of-view is centered on 20.5$\degree$ of Stonyhurst heliographic latitude and -2$\degree$ of Stonyhurst heliographic longitude. The black contour delimits the area of the pulsations detected in the 335~\AA~passband of AIA \citep[see Figure~4 in][]{froment2015}, for a sequence of data between June 05, 2012 11:14~UT and June 08, 2012 11:16~UT. Intensity pulsations with a period of 9.0~hr are detected in the large loops seen in the background image with a confidence level greater than 99$\%$. The different colors of the field lines show different loop bundles/regions in the field-of-view. The red lines correspond to the ones matching the detected pulsations. The white arrow indicates the line chosen as an input for the 1D loop simulation in Section~\ref{sec:simu}.}
                 \label{fig:carte_extrapol}
	\end{figure*}
	

We extrapolate the magnetic field of NOAA~AR~11499 from the photosphere to coronal altitudes using a LFFF model with the method described in \citet{nakagawa1972} and \citet{alissandrakis1981}. Most of the loops in the region studied are located in a plage region, with relatively weak and small-scale fields, so vector magnetograms are probably not reliable enough to use directly with a non-linear force-free field (NLFFF) model, as discussed in \citet{zhao_2016}.
Moreover, this active region does not show a sigmoidal shape, so that the differences between magnetic field extrapolations LFFF or NLFFF ought to be relatively small.

We use magnetograms from the Helioseismic and Magnetic Imager \citep[HMI;][]{scherrer2012} on board SDO. The HMI data (hmi.M\_720s series) are read with the routine \texttt{read\_sdo} from the Interactive Data Language SolarSoftware library. We choose the HMI magnetogram closest to the middle date of the AIA sequence, i.e. June 06, 2012 23:12 UT, close to the central meridian. As the magnetograms do not show large structural differences during the whole sequence, the use of one particular magnetogram for the analysis ought to be sufficiently accurate. In addition, since the date chosen is close to the passage of the active region at the central meridian, it allows us to minimize distortion effects and thus to compare the extrapolated field lines with the detected contour of excess Fourier power. The magnetogram and the background coaligned AIA image at 171~\AA~(see Figure~\ref{fig:carte_extrapol}) are $4 \times 4$ pixels binned to increase the signal-to-noise ratio.
The field-of-view (FOV) for the geometrical study is an area of $519 \times 444$~Mm$^{2}$. This FOV is centered on 20.5$\degree$ of Stonyhurst heliographic latitude and -2$\degree$ of Stonyhurst heliographic longitude. The calculation is made in Cartesian coordinates.

\medskip

Placed at  $z=0$, the magnetogram defines the bottom of the extrapolation box, which is tangent to the photosphere at the center of the FOV.
The magnetogram is padded with zeroes at each side of the two horizontal directions to minimize aliasing. The total size of the bottom of the box is then $750 \times 750$~Mm$^{2}$. There are periodic boundary conditions on the four sides of the $(x, y)$ domain.
We use a uniform grid with $ 2048 \times 2048$ points for the Fast Fourier Transform (FFT). The field lines are extrapolated up to an altitude of 300 Mm. 
For the purpose of the analysis, the resulting magnetic field was stored on a nonuniform grid centered on the main active region of the field of view with 350 points in the $z$ direction using intervals as small as 0.3 Mm near $z=0$. 
The force-free parameter, $\alpha$, is chosen to visually fit the visible loops in the whole FOV.
By an iterative method, we chose $\alpha = -5.4 \times 10^{-3}$ Mm$^{-1}$. 

\medskip

Some extrapolated lines are presented in Figure~\ref{fig:carte_extrapol}, including those colored in red that correspond to the pulsating loop bundle.  The red lines are selected with the criterion that at least 20\% of the line length should be enclosed by the contour of detected pulsations. We use the detected contour of pulsations at 335~\AA, the channel with the strongest Fourier signal \citep[][see Figure 4]{froment2015}, while the background image is at 171~\AA~because the loops are more discernable in this passband.

The correspondence between the red lines and the contour of detected pulsations is better for the eastern footpoints than the western footpoints. However, since the contour of detected pulsations was constructed using three days of data after remapping in heliographic coordinates \citep{froment2015}, we should bear in mind that geometric distortions are significant. 
 
\subsection{Simulation setup}
The 1D hydrodynamic model used is described in \citet[Section 2]{mikic2013}. The equations for mass, momentum and energy conservation take into account the loop expansion via the cross-sectional area, $A(s) \sim 1/B(s)$, where $s$ is the coordinate along the loop and $B(s)$ the magnitude of the magnetic field along the loop. 
$B(s)$ is specified from the values obtained from the LFFF extrapolation. The heating term is imposed \textit{via} a function $H(s)$ constant in time.

The parallel thermal conductivity is chosen to follow the classical Spitzer conductivity value. The radiative loss function is derived from the CHIANTI atomic database (version 7.1) \citep{dere1997,landi2013} with coronal abundances \citep{feldman_potential_1992,feldman_elemental_1992, grevesse1998, landi2002}. 
We assume that the plasma is totally ionized, only composed of protons and electrons and that the temperature of protons and electrons are equal: $n=n_e=n_p$ and $T=T_e=T_p$. 

Even though an option to artificially broaden the transition region at low temperatures is typically used in this code, as described by \citet{lionello2009} and \citet{mikic2013}, we do not employ this technique in the present study.  Since we analyze only a single loop, we can afford to use a very high resolution mesh to resolve the very steep gradients in the transition region, and run the code for a long time.  Therefore, we keep the true Spitzer thermal conductivity and radiative loss law without modification.  Future studies involving parameter scans over many loop solutions will likely require the artificial broadening of the transition region.

The initial condition is defined by a state of hydrostatic equilibrium with a parabolic temperature profile along the loop (2~MK at the mid-point of the loop) and a cool ($T_e=T_{ch} = 0.02$~MK) chromospheric reservoir at the footpoints. In this initial state, the chromosphere thickness is 3.5~Mm at each footpoint, which evolves for $t>0$ given the loop geometry and the heating function chosen. 
$T_{ch}$ and $n_{ch}$ are the boundary conditions at $s=0$ and $s=L$, with $n_{ch}=6 \times 10^{18}$~m$^{-3}$.


	\begin{figure}
		\centering
                 \includegraphics[width=\linewidth]{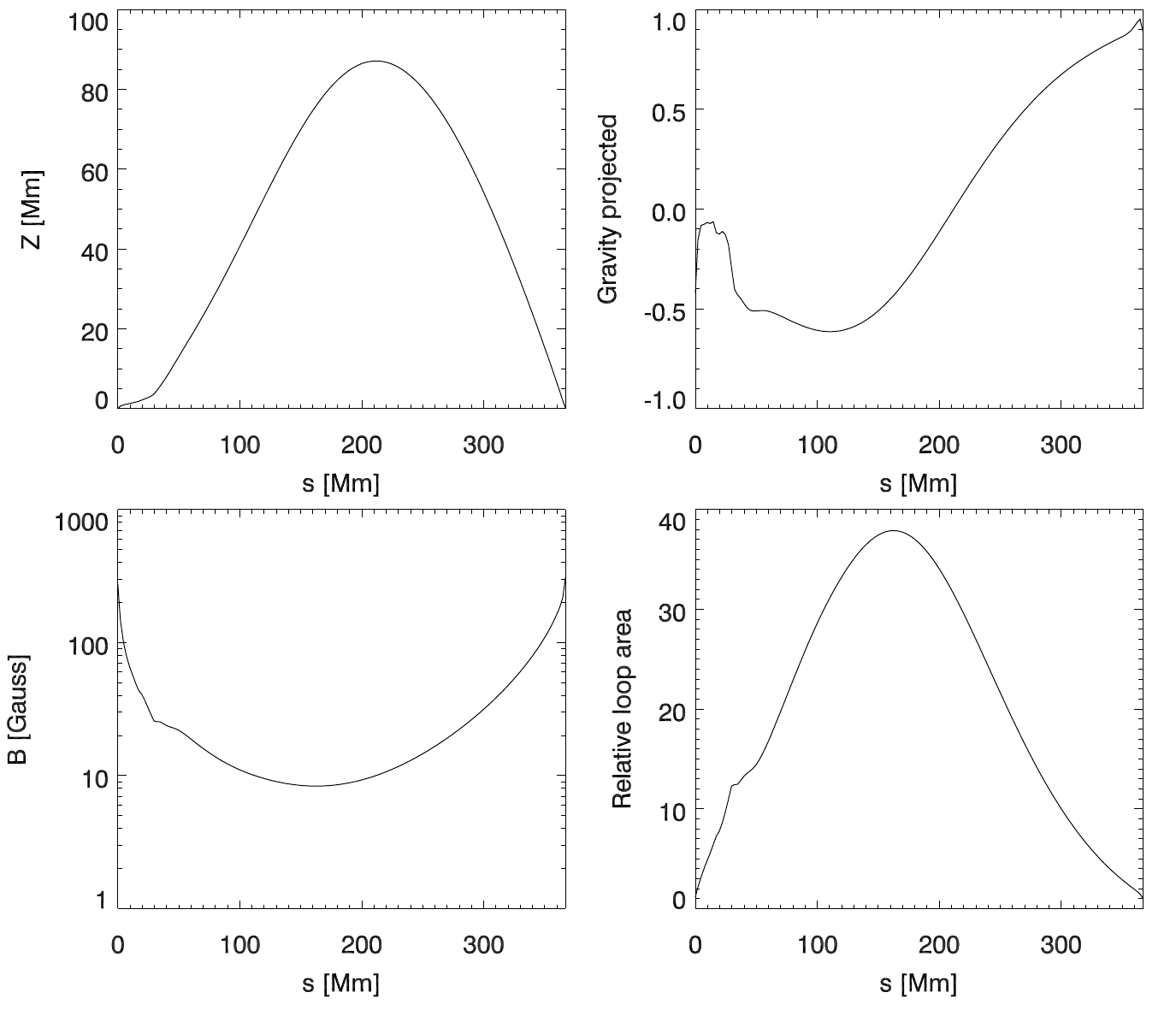}
                 \caption{Loop simulation inputs as a function of the position along the loop for the 1D hydrodynamic simulation presented in this paper. This geometry is from the extrapolated field line indicated by a white arrow in Figure~\ref{fig:carte_extrapol}. Top left: loop profile, altitude in Mm of each point along the loop. Top right: the acceleration of the gravity projected along the loop, normalized as presented in Equation~\ref{eq:gravity_projected}. Bottom left: the strength of the magnetic field $B(s)$ in Gauss. Bottom right: the loop expansion given by the evolution of the cross-sectional area A(s), normalized at the first footpoint.}
                 \label{fig:geometry_loop_simu}
	\end{figure}


\medskip

The simulated loop uses the geometry of the magnetic field line indicated by the white arrow in Figure~\ref{fig:carte_extrapol}. This field line is in the domain of loops with EUV intensity pulsations and has a length $L=367$~Mm. The geometry of this line is shown in Figure~\ref{fig:geometry_loop_simu}. 
Here we choose to conduct the 1D modeling using only one loop geometry. However, we have to bear in mind that we aim to model here the average behavior of the pulsating loop bundle observed with AIA. This simulation would therefore not be considered to be able to reproduce the details of what could happen more locally.

As input to the simulation, we give the loop profile, i.e. the altitude of each point of the loop, the gravity projected along the loop, and the magnetic field strength along the loop. The loop expansion is given by the cross-sectional area $A(s)$, obtained from $B(s)$, and normalized to its value at the first footpoint.
For this quite asymmetric loop, the apex is at an altitude of 87~Mm at $s=212$~Mm (i.e. at $0.58 \, L$), i. e. the loop is skewed towards one footpoint.
But the highest factor of asymmetry in this loop is the different behavior around the two footpoints. It is worth noting the shallow slope of the loop around the eastern footpoint ($s=0$). 
This behavior can be seen in the normalized gravity projected along the loop, defined as 
\begin{equation}
	\frac{\vec{g}\cdot\vec{B}}{\Vert\vec{g}\Vert \, \Vert\vec{B}\Vert}
	\label{eq:gravity_projected}
\end{equation}
with $\vec{g}$ the acceleration of the gravity.
While the value of the projected gravity is close to 1.0 (the field line is close to the vertical) at the western footpoint ($s=L$), it remains under 0.6 (in absolute value) in the eastern leg and drops around the footpoint to reach about 0.1. 

The magnetic field strength is about the same at $s=0$ ($\sim315$~Gauss), where $B_{z0}>0$, as at $s=L$  ($\sim275$~Gauss), where $B_{z0}<0$.  
The cross-sectional area at each footpoint is thus similar.
The evolution of the loop cross-sectional area, normalized to its value at $s=0$, shows that it is highly non-constant. The loop expansion factor reaches a value of 38 at $s=162$~Mm, i.e. before the loop apex. 

To solve the hydrodynamic equations, we use a nonuniform spatial mesh, fixed in time. Our run uses $100, 000$ mesh points. The mesh spacing is $\Delta s=2$~km in the chromosphere and transition region, increasing to $\Delta s=20$~km in the corona. The geometry profile taken from the extrapolations is linearly interpolated at the grid points. 
We verified that the fine mesh spacing region at low altitudes in the loop captures the gradients and the movements of the transition region.


	\begin{figure}
		\centering
                 \includegraphics[width=\linewidth]{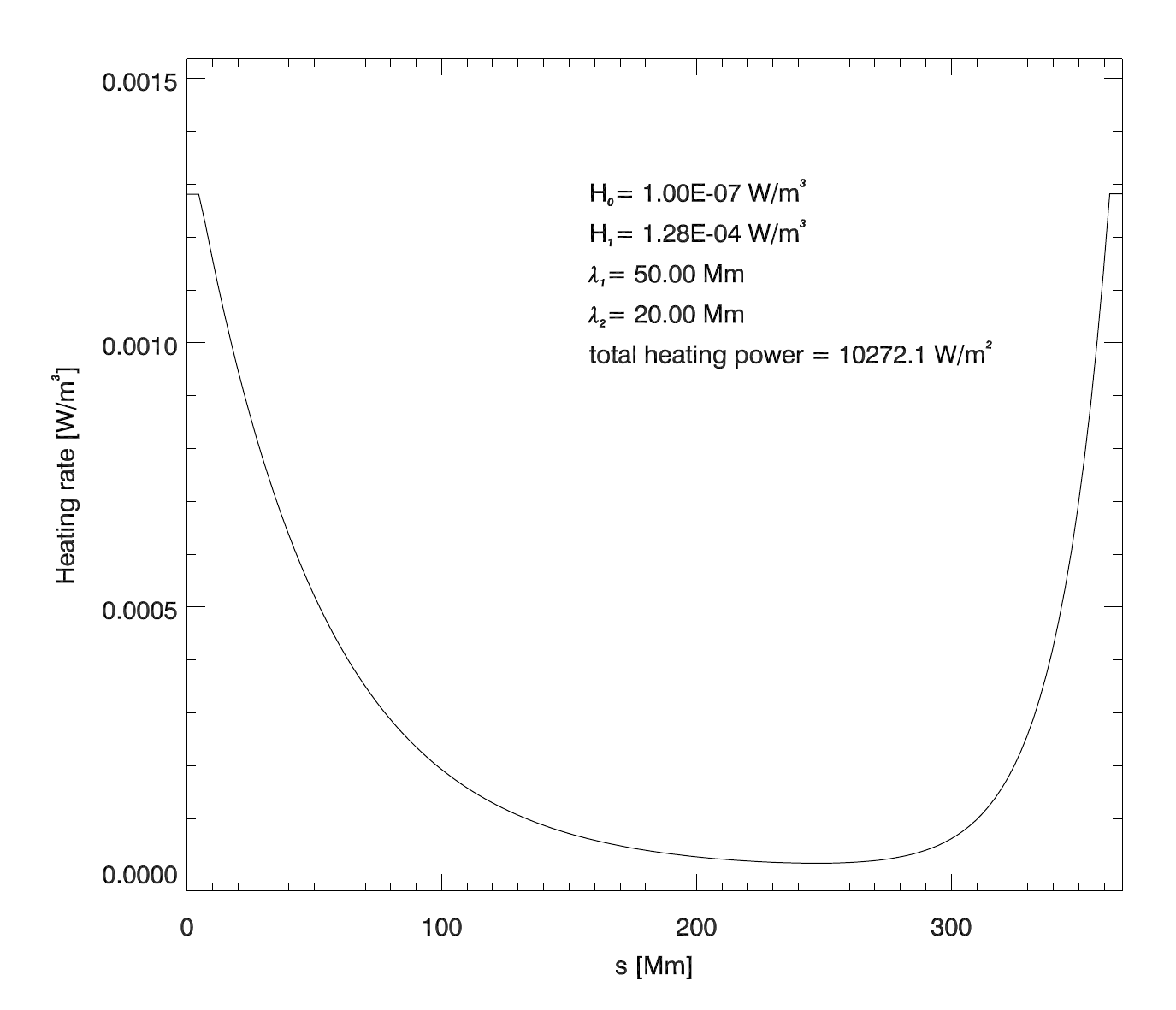}
                 \caption{Volumetric heating profile used for the loop simulation (Equation~\ref{eq:heating_model}).}
                 \label{fig:heating_profile}
	\end{figure}
	
	
\medskip
	
The heating function used for the simulation is similar to the one from \citet[Equation 6]{mikic2013}. This kind of heating function allows us to try different scale heights at each footpoint, to represent different possible stratifications and asymmetries of the heating:
\begin{equation}
H(s) = H_0 +  H_1(e^{-g(s)/\lambda_{1}}+e^{-g(L-s)/\lambda_{2}})   
\label{eq:heating_model}
\end{equation}

where $g(s)=\mathrm{max}(s-\Delta,0)$  and $\Delta=5$~Mm is the thickness of the chromosphere, where the heating is constant.

$H(s)$ is the volumetric heating rate, expressed in $\mathrm{W~m}^{-3}$. 
$H_0$ is the value to which $H(s)$ tends at the apex \footnote{only if $\lambda_{1,2} \ll \frac{L}{2}-\Delta$} and $(H_0+2\,H_1)$ is the value of the heating in the chromosphere.
$\lambda_1$ and $\lambda_2$ are the scale lengths for the energy deposition at the eastern and western leg of the loop, respectively.

We tested several heating configurations using equation~\ref{eq:heating_model} (see Section~\ref{sec:paperb}). Among the simulations showing TNE cycles we chose one that allowed us to produce a loop whose thermal behavior, as presented in Section~\ref{simu_thermal_behavior}, matches with the average loop bundle behavior observed with AIA.
We plotted the heating profile used for this simulation in Figure~\ref{fig:heating_profile}, with the following parameters: $H_0=1 \times 10^{-7} \, \mathrm{W~m}^{-3}$, $H_1=1.28 \times 10^{-4} \, \mathrm{W~m}^{-3}$, $\lambda_1=50 $~Mm and $\lambda_2=20$~Mm. The total heating power integrated over the loop length is about $10^5 \, \mathrm{W/m}^2$ and it is two times larger in the eastern leg ($0  < s < 212 \, \mathrm{Mm}$) than in the western leg ($212\, \mathrm{Mm} < s < L $). 
Most of the heating is concentrated around the loop footpoints since we chose $H_0 \ll H_1$ and the profile is asymmetric. 

\subsection{Conditions of occurence of TNE for this loop geometry}\label{sec:paperb}

In order to best reproduce the observed behavior of the pulsating loop bundle, i.e., the period of pulsation and the thermal evolution deduced from the DEM and the time lag analysis of \citet{froment2015}, we tested several scale lengths and strengths for $H(s)$. This scan of different heating conditions for one loop geometry allowed us to scan different loop behaviors: static loops as well as loops with evaporation and condensation cycles, whether incomplete or complete. For our particular loop geometry, these TNE cycles appear only when the heating is stronger at the eastern leg than at the western leg. This parameter space study will be presented in a forthcoming companion paper.

\subsection{Simulation results : thermal evolution of the loop}\label{simu_thermal_behavior}


	\begin{figure*}
		\centering
		
		\gridline{\fig{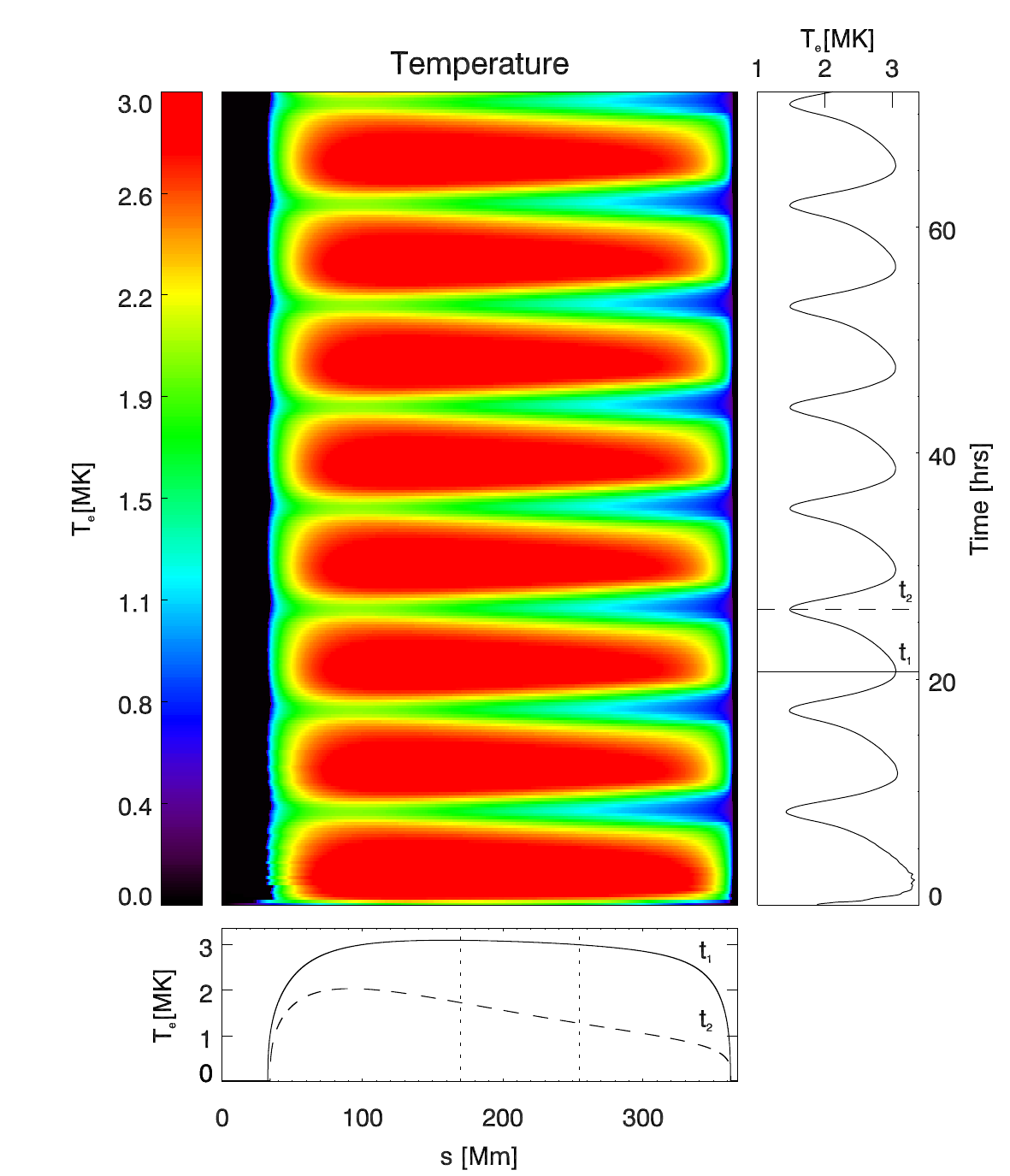}{0.45\textwidth}{}
         		     \fig{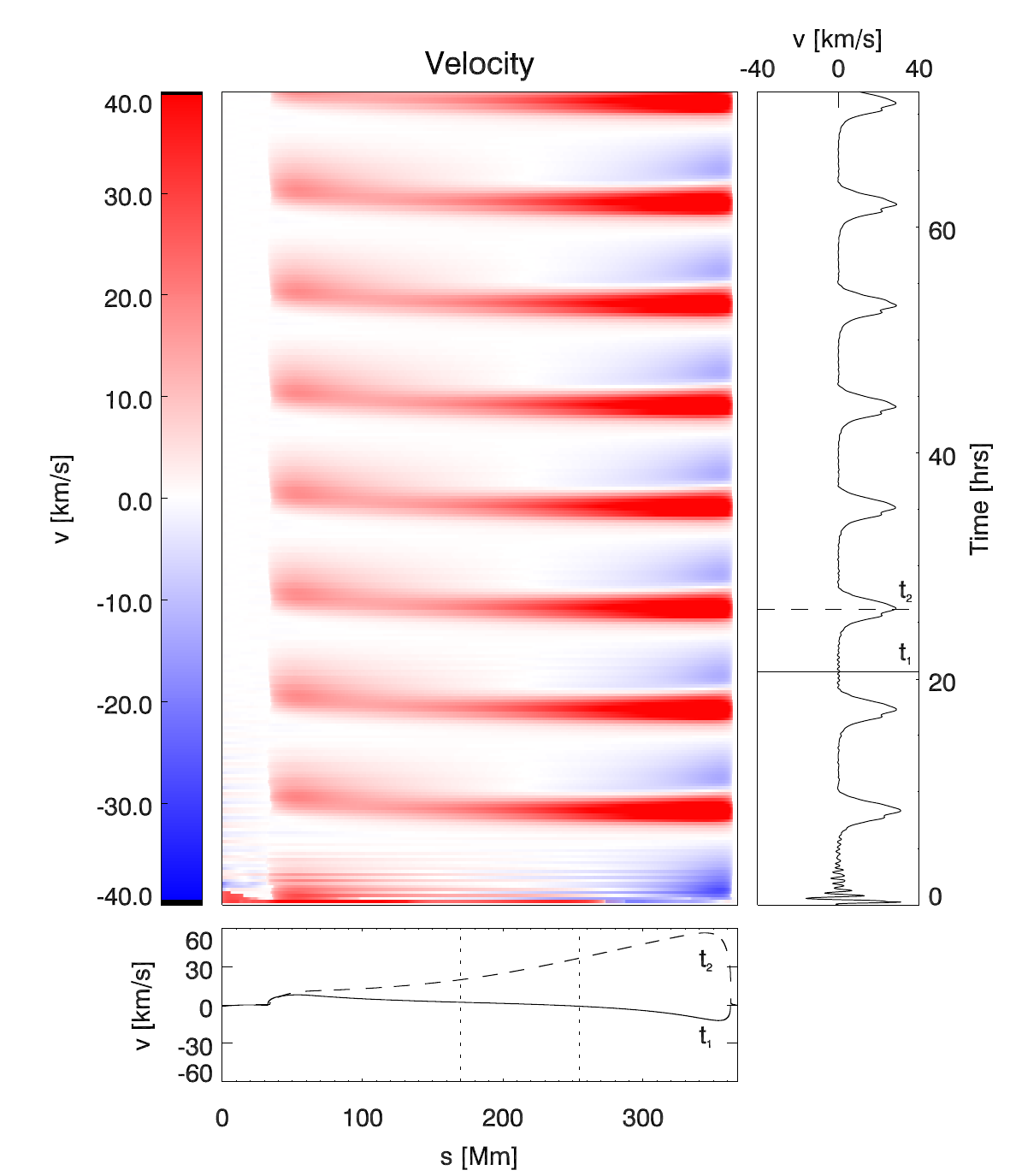}{0.45\textwidth}{}
			          }
		\gridline{\fig{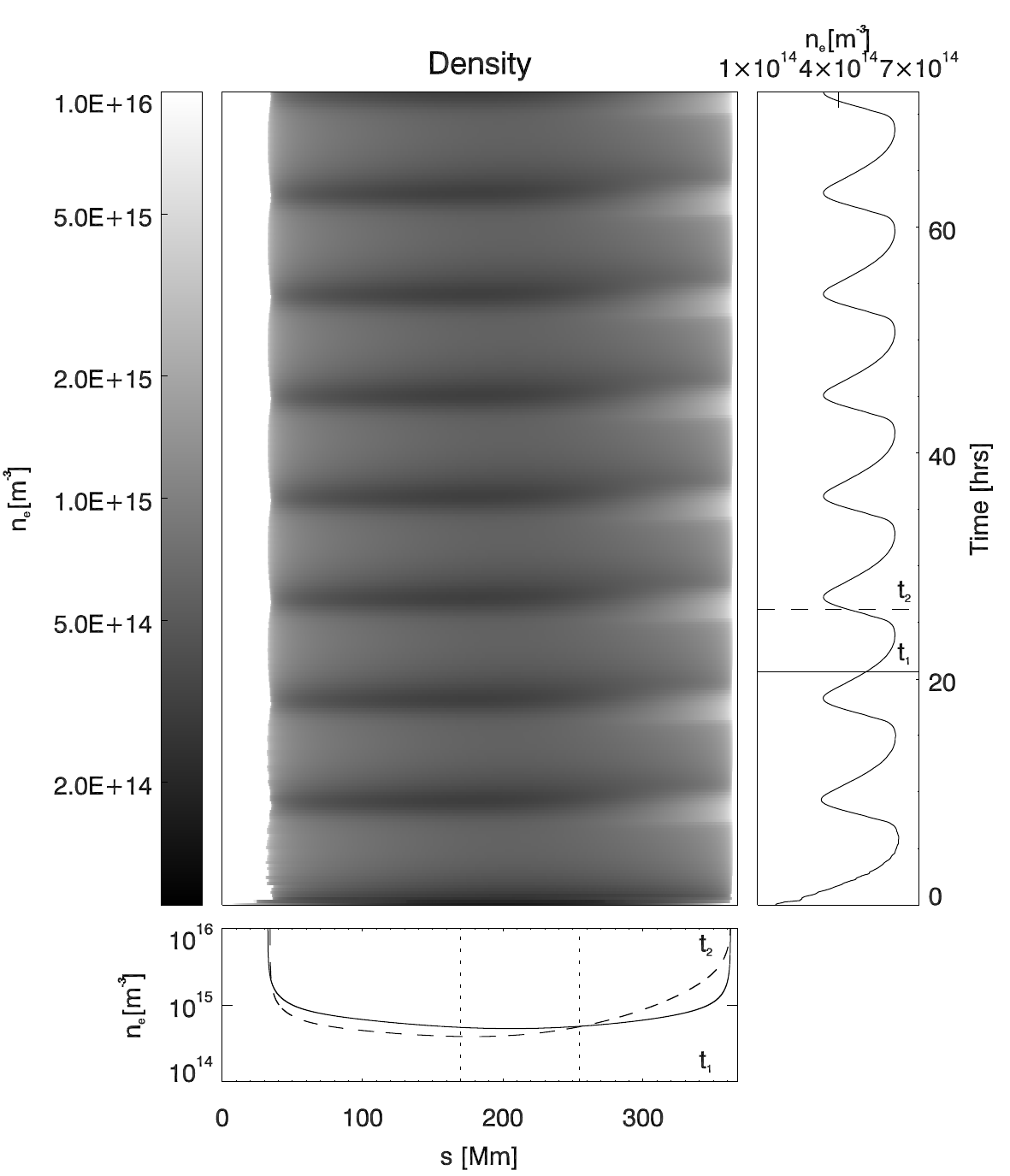}{0.45\textwidth}{}
         		     \fig{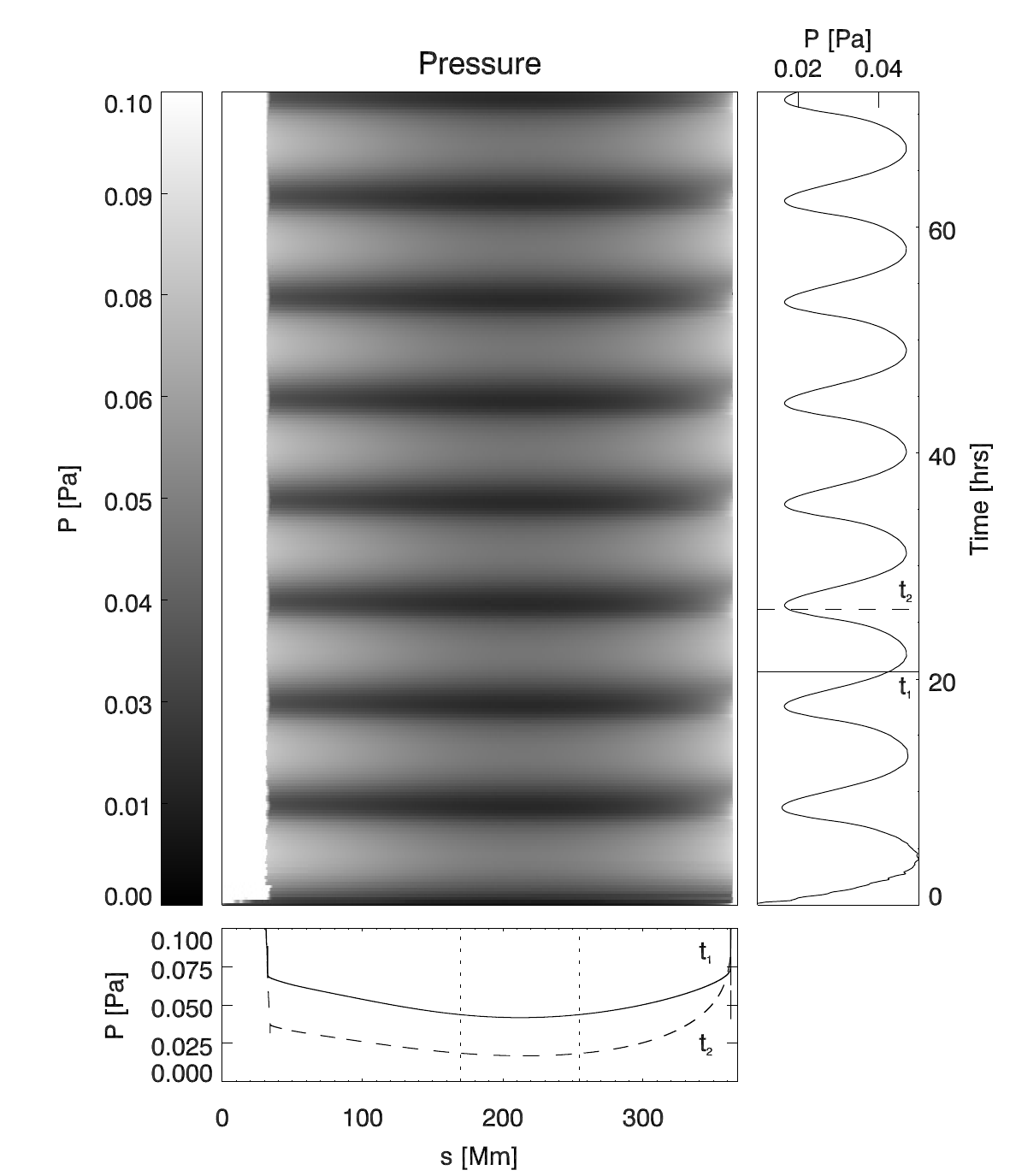}{0.45\textwidth}{}
			          }	          

  	\caption{Evolution of the temperature $T_e$, density $n_e$, velocity $v$,  and pressure $p$ for the loop simulated with the 1D hydrodynamic code \citep{mikic2013}. This loop is based on the geometry of the extrapolated magnetic field line designated by the white arrow in Figure~\ref{fig:carte_extrapol}. The constant heating function of Figure~\ref{fig:heating_profile} is applied. Top left: evolution of the temperature along the loop and during the 72 hr of the simulation. On the right of the 2D plot, we display the evolution of the temperature around the loop apex (mean value between the two dotted bars in the bottom panel). On the bottom of the 2D plot, we show two temperature profiles (solid and dashed lines, corresponding respectively to the hot phase at $t_1$ and the cool phase at $t_2$). The corresponding instants are indicated respectively by the solid and dashed lines on the plot of the temperature evolution around the apex. We plot the other loop physical parameters in the same way. Top right:  longitudinal velocity. Red (positive) for flows from the eastern footpoint to the western one and opposite for blue. Bottom left: density. On the 2D plot and the loop profiles, $n_e$ is shown in logarithmic scale. On the apex time series $n_e$ is in linear scale. Bottom right: pressure.}
	\label{fig:phy_param_simu}
	\end{figure*}


We simulate the loop evolution during a time sequence of 72~hr and analyse the loop behavior in response to the heating profile applied. The evolution of the temperature, density, velocity, and pressure is represented in Figure~\ref{fig:phy_param_simu} as a function of time and distance along the loop. On the right hand side of each 2D plots, we show the time evolution of the corresponding parameter averaged over the loop apex region, defined as the part of the loop above 90~\% of the apex height. We mark it with the two dotted bars in the lower plot of each panel. The profiles in the bottom panels correspond to instants of the simulation, $t_1$ and $t_2$, marked respectively by the solid and dashed bars in right hand side plots.

\medskip

These profiles show that this loop experiences cyclic TNE pulsations with aborted condensations, i.e. the apex temperature never falls below 1~MK. These condensations are falling along the western leg which is heated about half as much as the eastern one. There are pulsations with a period of about 9~hr for each physical parameter. It is the same period as observed in the EUV intensities and DEM parameters by \citet{froment2015}. The 2D plots and the profiles show a large chromospheric part around the eastern footpoint. It is due to the geometry of the loop, i.e. the flattened eastern leg of the loop at low altitude. Apart from this peculiarity in the geometry, this case is similar to Case 10 presented in Figure~14 of \citet{mikic2013}. The cycles seen for all the physical parameters are nearly identical over the duration of the simulation. 

This loop is relatively hot as the coronal part has a peak temperature of 3~MK. The temperature always remains above 1.4~MK for the eastern part of the loop and around the loop apex. It drops to 0.6~MK in the western leg at the end of the cooling phase, when the condensations are the strongest. Most of the loop remains in coronal conditions at all times.

During the rise of the temperature in the loop, there are upflows at both the eastern and the western footpoints with a velocity of about $10 \, \mathrm{km~s}^{-1}$.
During the cooling phase, the density decreases to $\sim 4 \times 10^{14}$ m$^{-3}$ around the apex and in the eastern part of the loop. These parts of the loop are drained. The density increases by a factor of 3 in the western leg due to condensations. 
As a consequence, the pressure drops everywhere in the loop with a minimum value around the apex.

Due to the drop in the pressure, siphon flows are seen from the eastern part to the west footpoints (velocity of about $60 \, \mathrm{km~s}^{-1}$) of the loops.
 \citet{mikic2013} hypothesize that these siphon flows play a key role in aborting the condensations. However, this remains to be confirmed as some 2.5D modeling results show that such siphon flows could actually provide more density to the condensations \citep{fang_coronal_2015}.

The evolution of the density around the apex is delayed compared to the evolution of the temperature, as expected for a state of TNE \citep[e.g.][]{mikic2013}, in conjunction with the periodic evolution of these two parameters. We notice that the delay between the maxima of temperature and density is larger than the one between their minima. This property was previously seen in \cite[see Figure~6]{mikic2013}.

\section{Do the properties of the simulated loop match the observations?}\label{sec:obs}

	\subsection{Potential evidence of asymmetric heating}

In Section~\ref{sec:simu}, we have traced magnetic field-lines from a LFFF extrapolation of event~1 of \citet{froment2015}. We presented a 1D hydrodynamic simulation of a loop with a geometry taken from one of the extrapolated field lines matching the contour of detected pulsations. This loop has a large low-lying portion at its eastern footpoint. 
Moreover, we found that we can reproduce the properties of the observed pulsating loops only if there is more heating at the footpoint with this particular geometry (see Figure~\ref{fig:heating_profile} and Section~\ref{sec:paperb}). The full parameter space study will be presented in a forthcoming companion paper.


	\begin{figure*}
		\centering
		\gridline{\fig{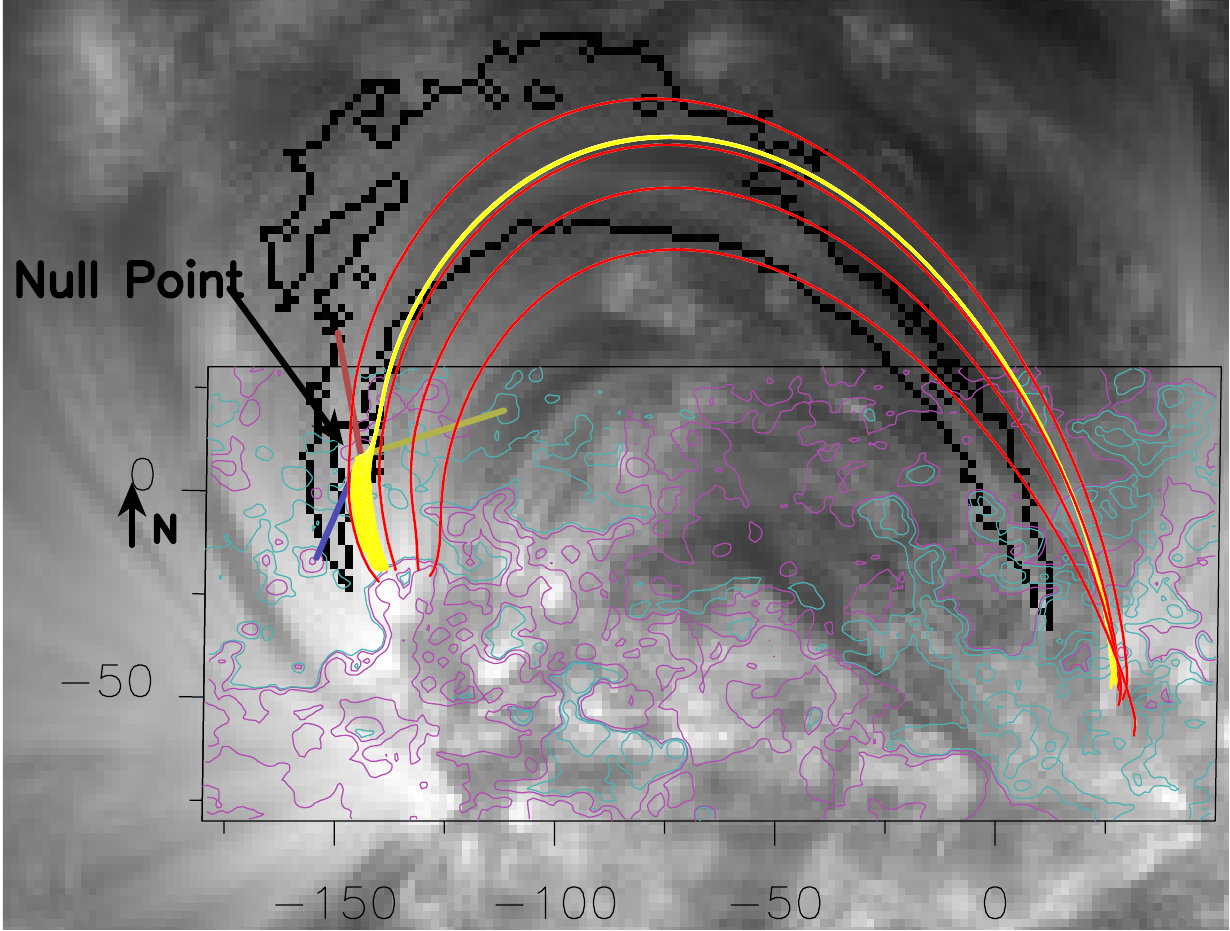}{0.4\textwidth}{}
         		     \fig{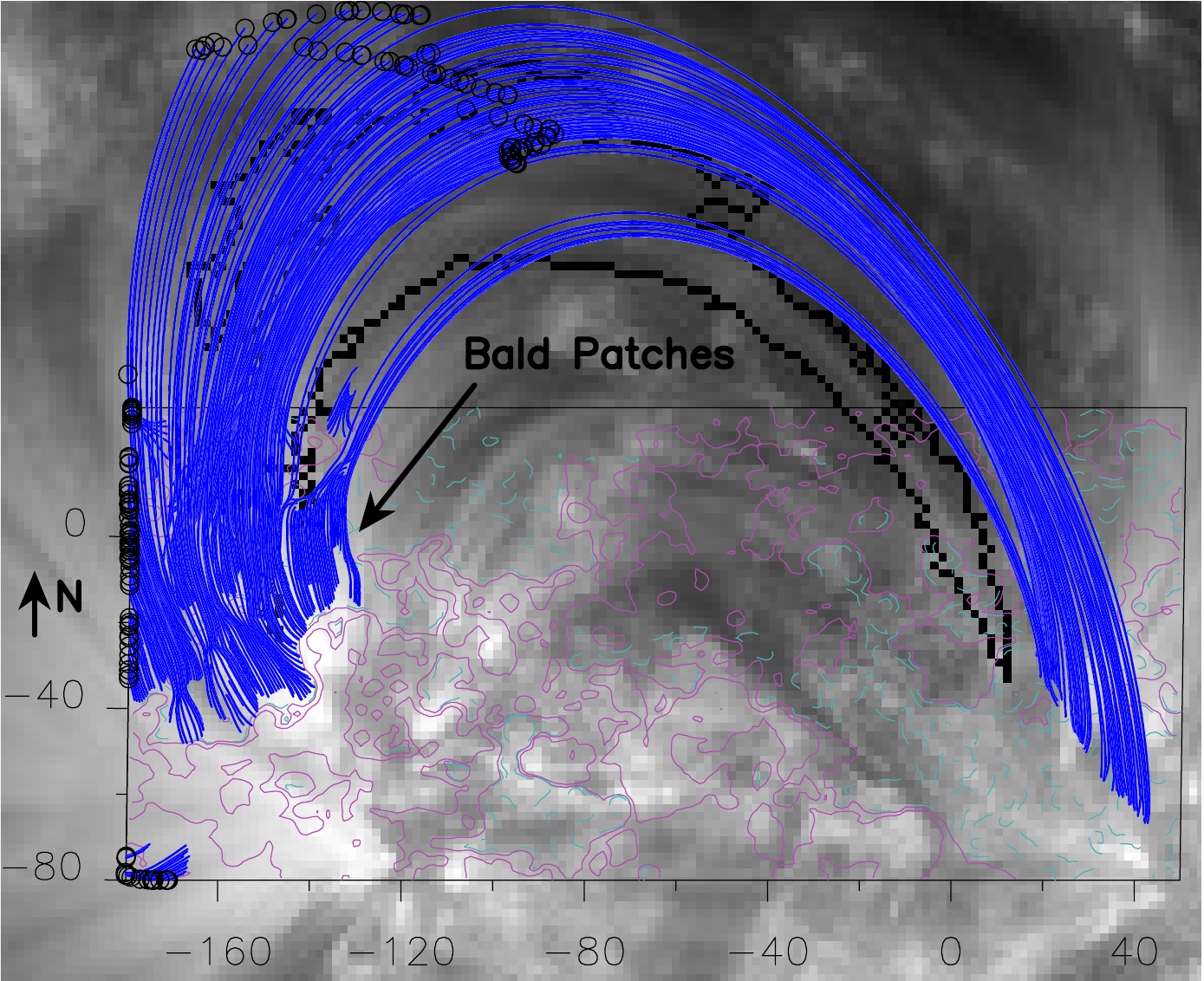}{0.4\textwidth}{}
			          }
		\gridline{\fig{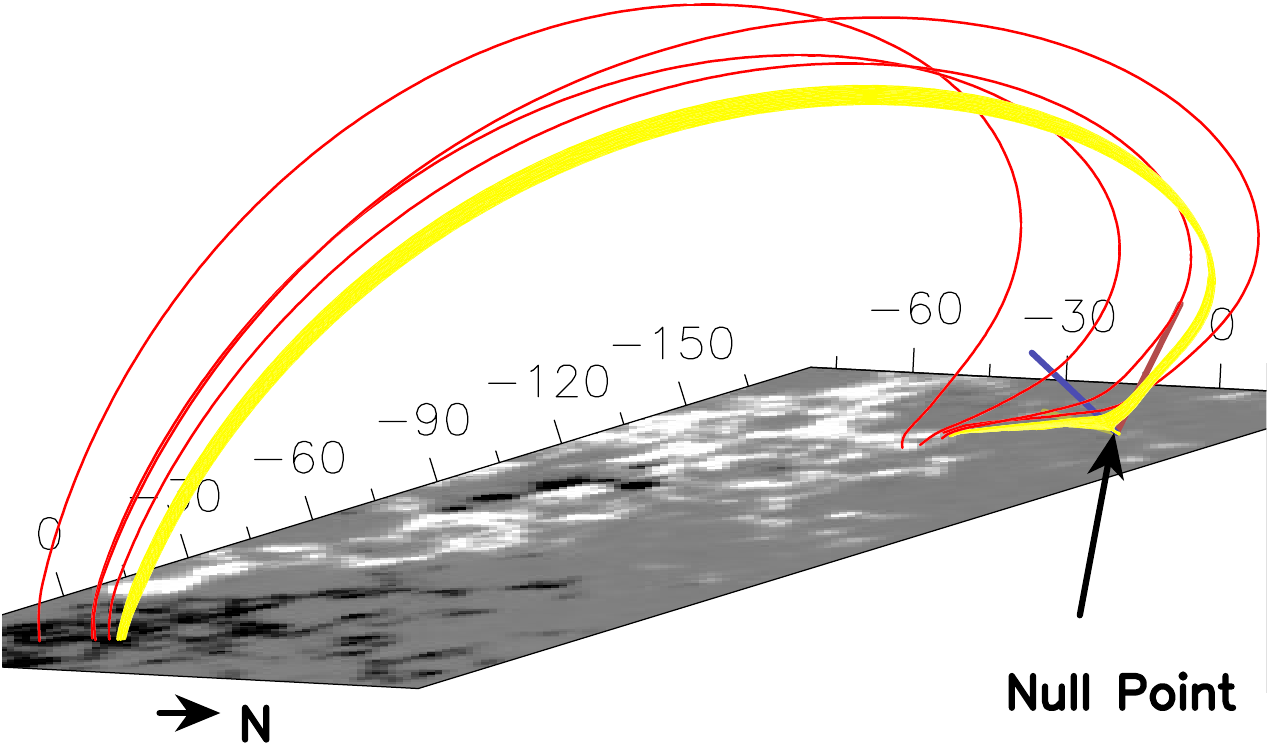}{0.4\textwidth}{}
         		     \fig{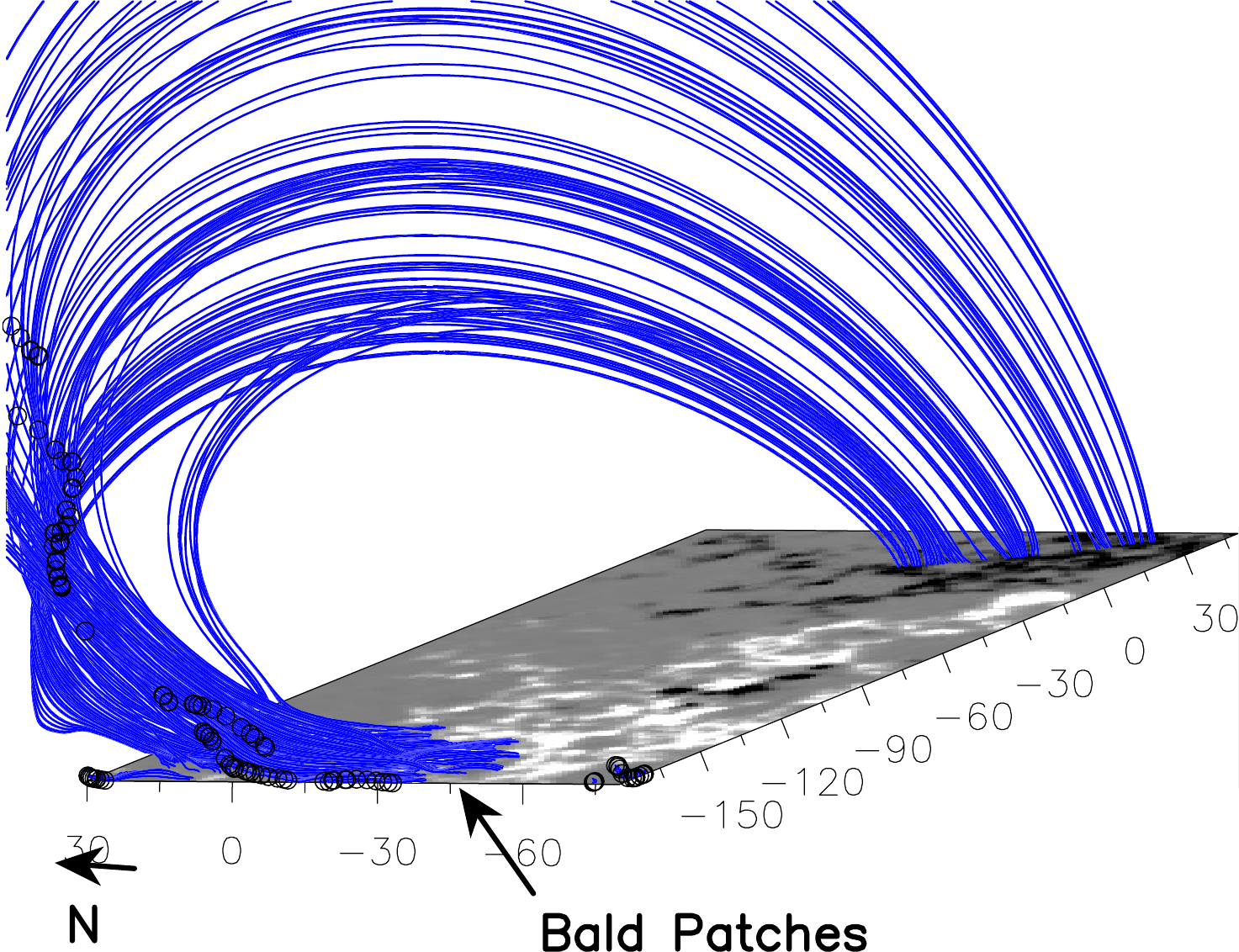}{0.4\textwidth}{}
			          }	   

		  \caption{Analysis of the magnetic topology of the region. Top and bottom left: AIA view (top) and side view (bottom) of the low-lying null point. Some magnetic field lines are traced in yellow around the null point, marked by the small coordinate system (red, yellow and blue bars). We add the same red lines as in Figure~\ref{fig:carte_extrapol}. Top and bottom right: AIA view (top) and side view (bottom) of field lines (in blue) with a bald patch topology on the eastern footpoints region. For the AIA view, the $\pm$10 G,  $\pm$100 G, and  $\pm$1000 G contours from the HMI magnetogram are displayed in purple for positive values and in cyan for negative ones. The black circles indicate the top of the field lines connecting outside of the box. Black arrows indicate the north direction.}
                 \label{fig:spe_loopfpt}
	\end{figure*}


	\begin{figure*}
		\centering
		\gridline{\fig{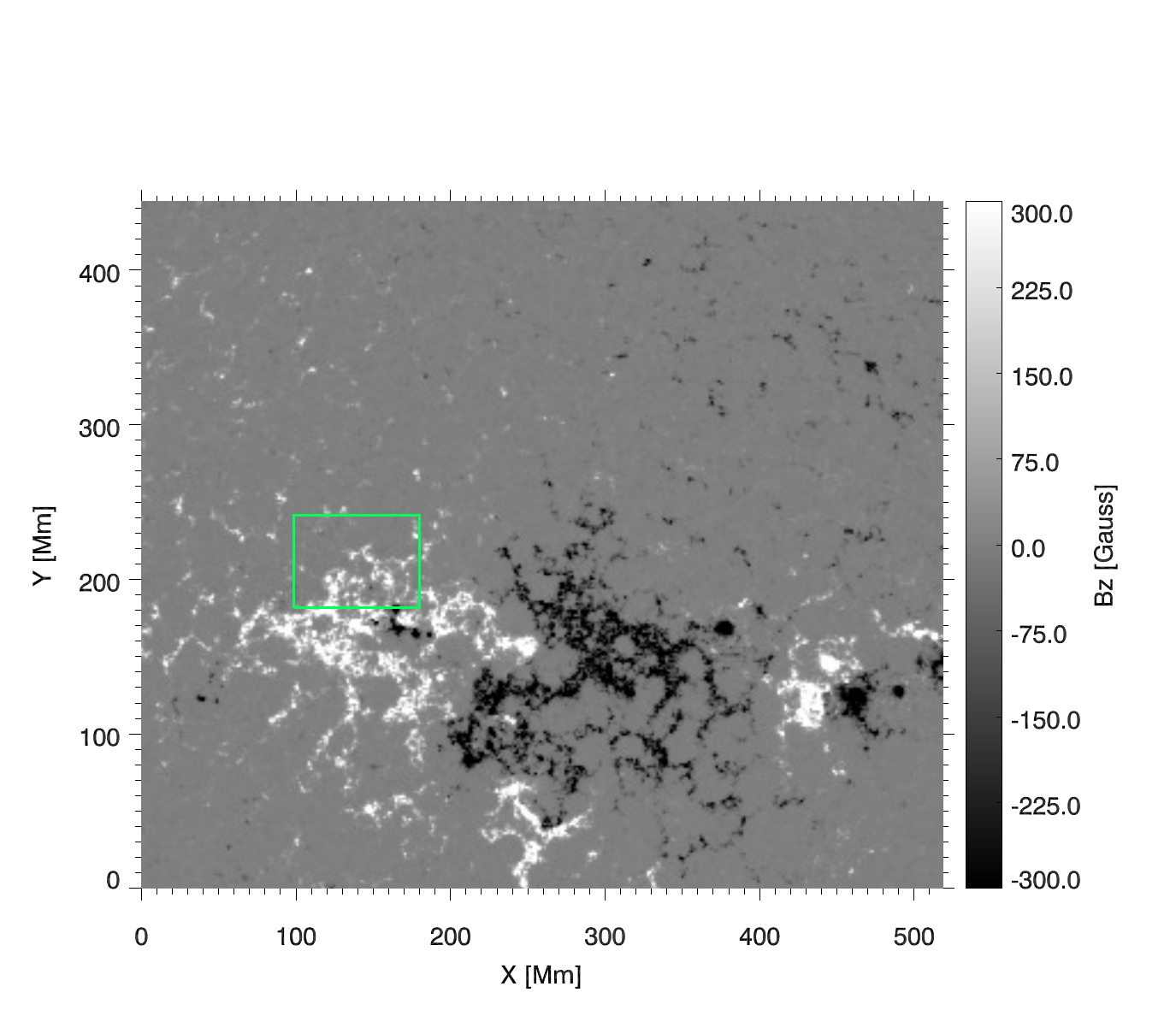}{0.5\textwidth}{}
			\fig{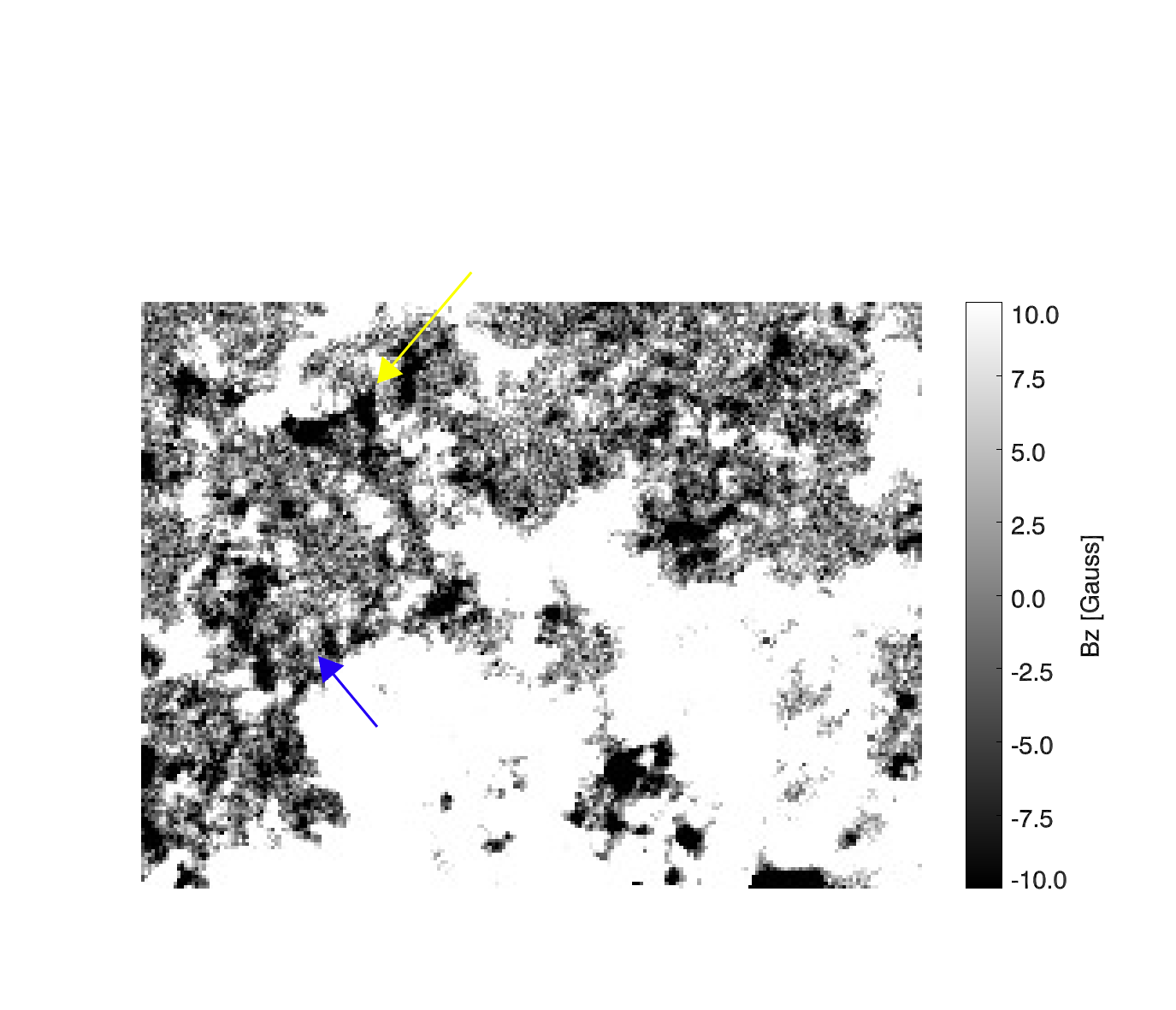}{0.5\textwidth}{}
			          }
			   
		  \caption{Left: The HMI magnetogram on June 06, 2012 23:12 UT. It covers the field of view used for the extrapolations (519 Mm $\times$ 444 Mm). The green box delimits the area where we zoom (right panel) to focus on topological elements. Right: Zoom on the magnetogram in the area where most of the field lines matching the pulsating loops are rooted. The blue and yellow arrows indicate, respectively, areas corresponding to the bald patches region and to the photospheric null point region. For this zoom, we use the non-binned HMI magnetogram version.}
                 \label{fig:spe_mag}
	\end{figure*}

\medskip

Examining the magnetic topology around the eastern pulsating loops footpoints, we found one photospheric null-point (at $z=370 \, \mathrm{km}$) and many bald patches. This particular magnetic topology explains the field line geometry in this area. In Figure~\ref{fig:spe_loopfpt}, we show an AIA view and a side view of the null point and bald patches. In the left panels, we trace in red the lines matching the pulsating loops, as in Figure~\ref{fig:carte_extrapol}. In the side view, we see that at least two field lines have a grazing leg (this includes the field-line used for the simulation), due to the presence of the null-point. In the right panels, we trace some field lines anchored in the bald patch area in the East of the region. These blue lines clearly show the different conditions at their two footpoints. As pointed out in Figure~\ref{fig:spe_mag}, we can see in the HMI magnetogram that the footpoints where we found the null-point and the bald patches are located on the external side of a plage region. On the magnetogram zoom (right panel), we can see that the bald patches are located in an area adjacent to the plage region with many alternations of small scale negative and positive polarities (pointed by the blue arrow in Figure~\ref{fig:spe_mag}). The low-lying null is located in a meso-scale area (alternation of positive, negative and positive polarity, pointed by the yellow arrow in Figure~\ref{fig:spe_mag}).

The specific magnetic topologies and the occurence of permanent photospheric motions observed in the magnetograms in this area could favor reconnection \citep[e.g.,][]{billinghurst_current-sheet_1993,pariat_current_2009}.
It thus might be evidence for different heating conditions at each footpoint of these loops, which might conceivably be the source of the asymmetric heating needed to drive the solution of the 1D simulation into TNE cycles.
Although we study here the magnetic topology of the region at only one date, the magnetograms show this area of alternation of small scale positive and negative polarities during the sequence of $\sim$~6 days. The potential site of reconnection could thus last for several days and produce quasi-steady favorable conditions for the appearance of pulsations. Moreover, we made the same extrapolations as presented in Section~\ref{sec:mag_extrapol} using a magnetogram at the beginning and at the end of the three-day sequence presented in \citet{froment2015} (between June 05, 2012 11:14 UT and June 08, 2012 11:16 UT). We checked that the overall geometry of the field lines matching the contour of detected pulsations (such as asymmetry) does not change significantly between these two dates and the one studied here.

Such a potential site of reconnection is likely also present at other locations in this active region. These heating conditions, as well as the geometry of the loops studied, are not generic properties of loops showing long-period intensity pulsations. However, as we will demonstrate in the companion paper, both the loop geometry and the heating conditions influence the production of TNE. The pulsating loops are not characterized by a single common property but rather by a combination of the right heating conditions and geometry.

\subsection{EUV intensities and time lags}\label{sec:intensity}
	
	\subsubsection{Synthetic intensities as seen in the AIA coronal channels}

	
	\begin{figure}  
		\centering
                 \includegraphics[width=\linewidth]{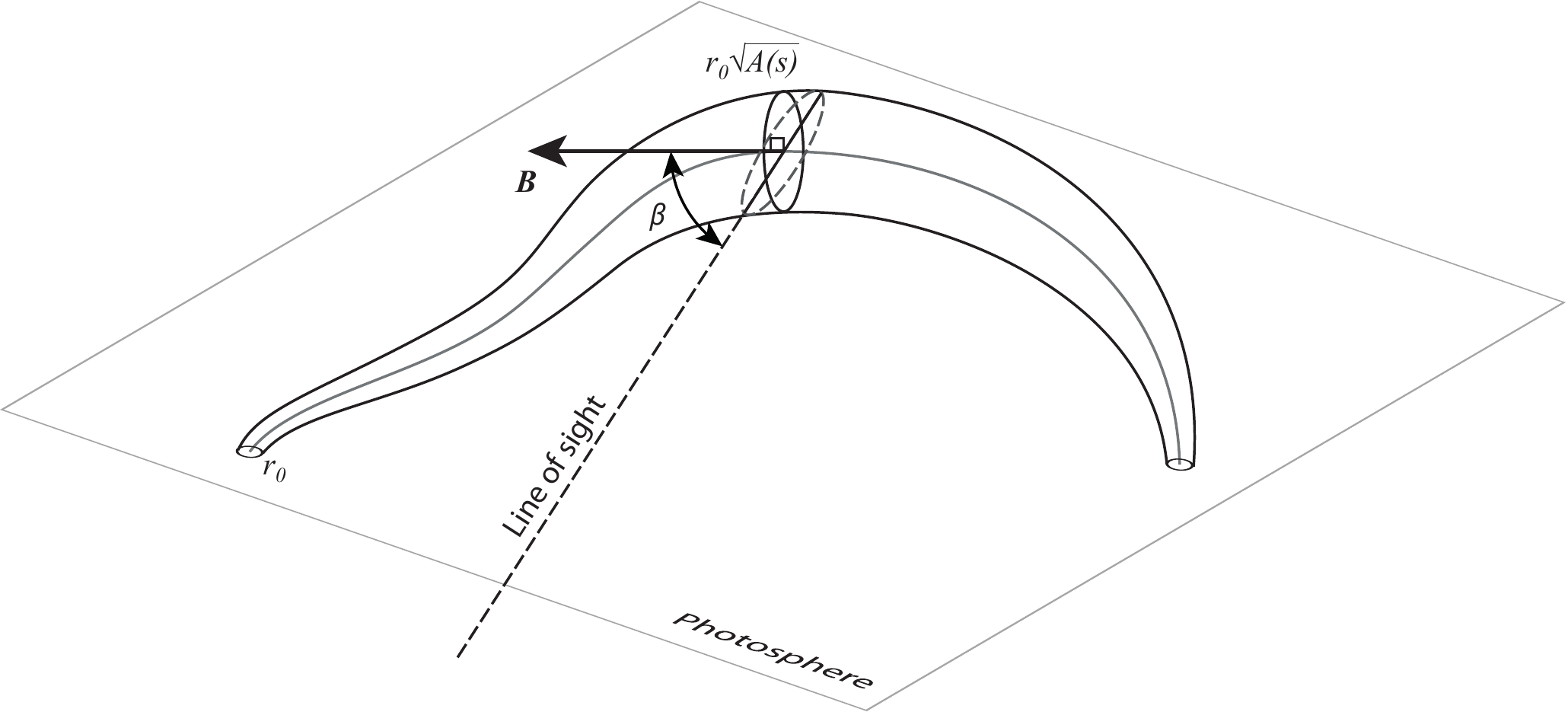}
                 \caption{Sketch of the projection of the loop cross-section thickness as seen from AIA along the line of sight corresponding to Figure~\ref{fig:carte_extrapol}. We use this projection for the calculation of the synthetic intensities (Equation~\ref{eq:synthetic_intensities_2}). $r_0$ and  $r_0 \sqrt{A(s)}$ are respectively the radius of the loop cross-section at the first footpoint and everywhere along the loop. $\beta$ is the angle between $\vec B$ and the direction of the light of sight (in dashed line outside the loop and in solid line in the loop). The loop cross-section is represented in black solid line at the loop apex, the projected loop cross-section is represented in grey dashed line.}
                 \label{fig:sketch_int}
	\end{figure}
		

We use the simulated temperatures and densities along the loop to produce synthetic intensities as they would be observed in the six coronal passbands of AIA, in the geometric configuration of Figure~\ref{fig:carte_extrapol}.
We produce these intensities from the AIA response functions $R_b(n_e,T_e)$ to isothermal plasma for each channel, calculated with CHIANTI version 8.0 \citep{delzanna2015} \footnote{We use here the latest version of CHIANTI instead of the one used for the radiative loss function for the simulation (version 7.1).The differences with the use of CHIANTI version 8.0 for the radiative loss function should be very small.}. 
To simulate the intensities as they would be seen in EUV images, we perform the integration along the line of sight in the geometry corresponding to that given in Figure~\ref{fig:carte_extrapol}:
\begin{equation}
I_{simu} = \frac{1}{4\pi} \int n_e^2 R_b(n_e,T_e) dl
\label{eq:synthetic_intensities}
\end{equation}

We assume that for the average loop bundle modeled, $n_e$, $T_e$ and $A(s)$ vary slowly from one section to another, which is valid for the coronal part of the loop. So that we can assume that $n_e$ an $T_e$ are constant in each cross section of the loop.
We use the projection of the extrapolated magnetic field line as seen by AIA (see Figure~\ref{fig:carte_extrapol}) to calculate the thickness of the loop. 
We thus approximate Equation~\ref{eq:synthetic_intensities} as:
\begin{equation}
I_{simu} \sim \frac{1}{2\pi} \frac{r_0 \sqrt{A(s)}}{\sin(\beta)} n_e^2 R_b(n_e,T_e) 
\label{eq:synthetic_intensities_2}
\end{equation}

$\beta$ is the angle between $\vec B$ and the direction of the line of sight, as illustrated in the sketch of Figure~\ref{fig:sketch_int}. 
The $\beta$ values along the loop are determined using the projection of $\vec B$ along the line of sight. The $\beta$ values are between $60\degree$ and $120\degree$ in most of the loop (except for the footpoints).
We arbitrarily set $r_0=100$~km, the radius of the loop cross-section at $s=0$. Given the loop expansion, the radius of the loop cross-section reaches a maximum of $\sim 0.6 $~Mm. With a higher $r_0$ this value would be higher indeed. However, this does not change the relative evolution of the intensities along the loop that we aim to study.

	\subsubsection{Analysis of the loop intensity profiles}
	
	
	\begin{figure}  
		\centering
                 \includegraphics[width=\linewidth]{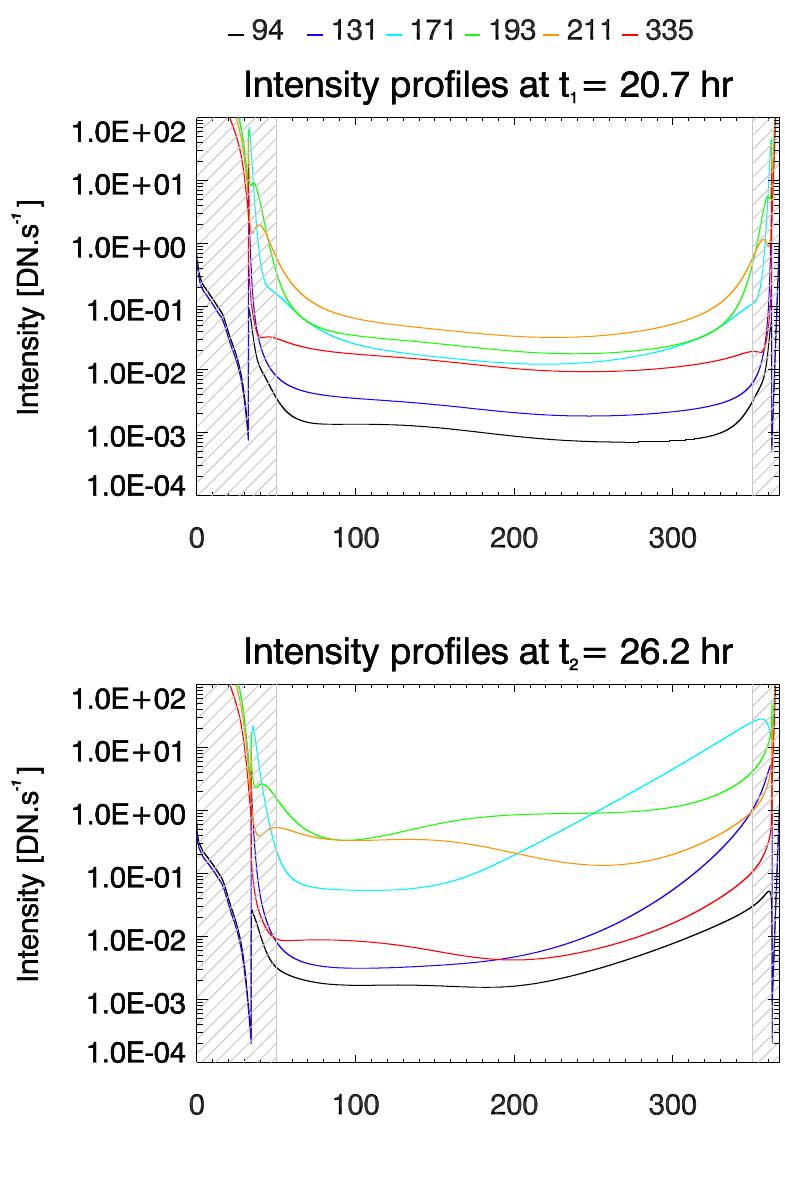}
                 \caption{Synthetic intensities along the loop for the six channels of AIA. Top: intensity profiles at $t_1 = $ 20.7~hr, that corresponds to the hot part of the cycle chosen. Bottom: intensity profiles at $t_2 = $ 26.2~hr, that corresponds to the cool phase of the cycle chosen. The parts of the loop profiles under $s=70$~Mm and above $s=350$~Mm, i.e. the grey hashed regions, are not considered.}
                 \label{fig:profile_int_loop_simu}
	\end{figure}
		

In Figure~\ref{fig:profile_int_loop_simu}, we trace the synthetic intensity loop profiles obtained for the six passbands. On the top panel, the intensity profiles correspond to the ones at $t_1=20.7$~hr in the simulation, when the temperature peaks. In the bottom panel, we plot profiles at $t_2=26.2$~hr, when the condensations are the strongest for this cycle.
The shape of the synthetic intensity along the loop should be examined only in the coronal part of the loop as the 1D code is not supposed to model properly the chromosphere. In addition, Equation~\ref{eq:synthetic_intensities} is only true in the case of an optically thin plasma, and around the loop footpoints the integration method of Equation~\ref{eq:synthetic_intensities_2} is approximate. We thus only consider loop profiles between $s=50$~Mm and $s=350$~Mm, i.e. outside the hatched regions.
At $t_1$, the intensity varies smoothly along the loop in all the channels and is not more than 10 times larger from one side to another. 
At $t_2$, the intensity values are on average larger in all the bands than at $t_1$ as the loop is cooling, especially for the bands with a cool temperature response wing  : 94~\AA, 131~\AA, 171~\AA, and 193~\AA. The 335~\AA~and 211~\AA~bands are less sensitive to these temperatures. Because of condensations in the western leg, the differences between the two legs are larger than at $t_1$, reaching $\sim 100$ at 131~\AA~and 171~\AA. However as the condensations are aborted, there are no sharp structures seen in the intensity profiles, in contrast to the simulations of \citet{klimchuk2010}. The large variation between the footpoints at $t_2$ (at 171~\AA), is consistent with the intensity variations visible at 171~\AA~(variations with a factor of about 80 for the pulsating loop bundle of Figure~\ref{fig:carte_extrapol}).

The 131~\AA~and 171~\AA~intensity variations between $t_1$ and $t_2$ at the western footpoint are large. However, we simulate here a single 1D loop. Due to line of sight integration of multiple structures, one can expect lower variations in observations.
We can conclude that these results are consistent with the conclusions from \citet{lionello2013} from 3D simulations: a loop experiencing TNE can have properties that match those of observed loops, with here the inherent limitations of 1D simulations (single 1D loop).

	
	\begin{figure*}  
		\centering
                 \includegraphics[width=\linewidth]{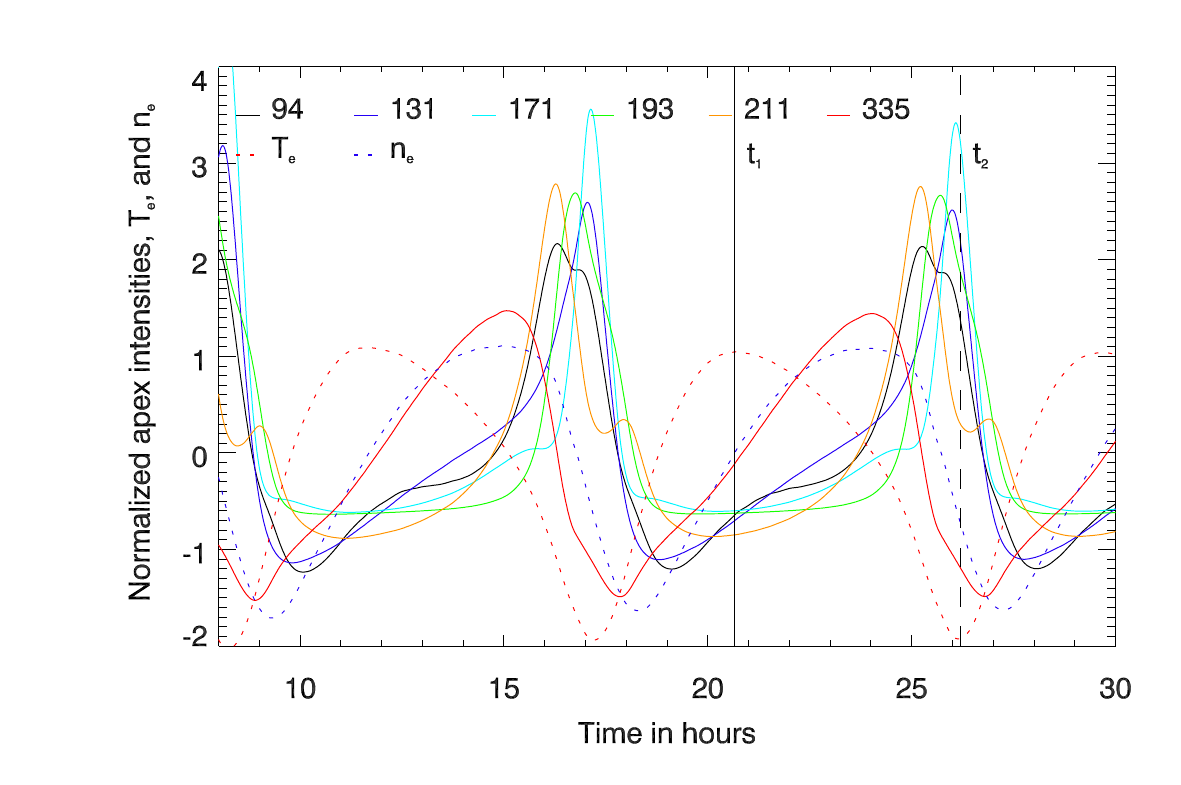}
                 \caption{Synthetic intensity time series in the six coronal channels of AIA in the geometry of Figure~\ref{fig:carte_extrapol}. For each channel, the intensity is averaged around the loop apex, i.e. between the two dotted bars indicated in the loop profiles of Figure~\ref{fig:phy_param_simu}. We zoom on two cycles present in the simulation, i.e. between 8 hr and 30 hr. In dotted lines, we overplot the evolution of the temperature $T_e$ (in red) and of the density $n_e$ (in blue), averaged around the loop apex. The intensities, $T_e$ and $n_e$ are all normalized to their standard deviation (we substract the mean curve and divide by their standard deviation). The solid bar and the dashed bar indicate respectively $t_1$ and $t_2$, the instants for which we plot the hot and the cool profiles in Figure~\ref{fig:phy_param_simu} and Figure~\ref{fig:profile_int_loop_simu}.}
                 \label{fig:int_loop_simu}
	\end{figure*}
		

	\subsubsection{Analysis of the time lag signature of cooling}

We trace the six time series of the synthetic intensities, averaged around the apex and normalized to their standard deviation (we subtract the mean curve and divide by their standard deviation) in Figure~\ref{fig:int_loop_simu}. We look at the evolution between 8 hr and 30 hr after the beginning of the simulation, i.e. approximately two cycles of the simulation. The synthetic intensity peaks first in the hotter channels and then in the cooler channels with the following order: 335~\AA, 211~\AA, 94~\AA~(considering the main peak of the response function), 193~\AA, 131~\AA, and 171~\AA. Since 335~\AA~peaks first, the plasma does not reach the temperature of the hotter peaks of the 94~\AA~and 131~\AA~channels. 
	

	\begin{figure}
		\centering
                 \includegraphics[width=\linewidth]{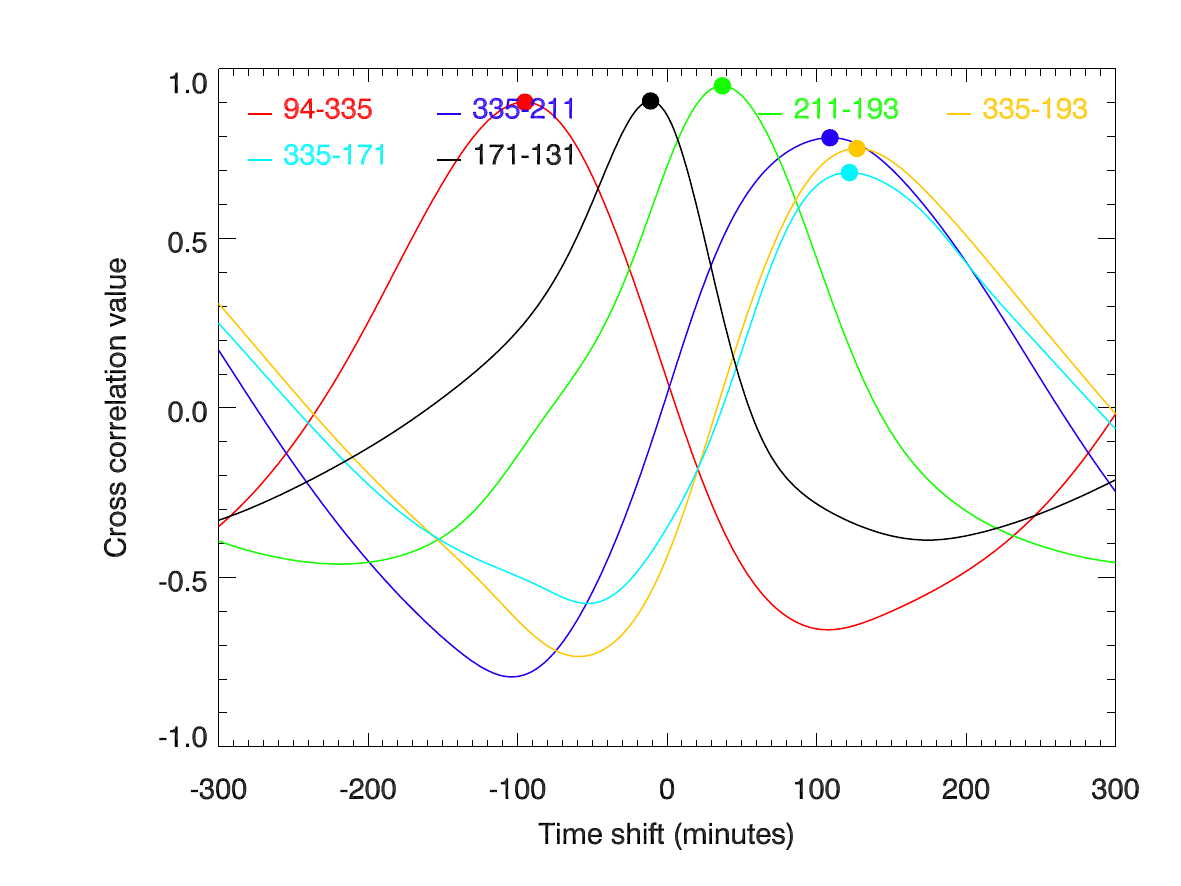}
                 \caption{Time lag analysis for the loop simulated in Section~\ref{sec:simu}. The cross-correlation values are given for six pairs of synthetic AIA intensities, averaged over loop apex: 94-335 (red), 335-211 (blue), 211-193 (green), 335-193 (orange), 335-171 (cyan) and 171-131(black). We explored time shifts from -300 minutes to 300 minutes. The time lag for each pair of channels is indicated by a colored dot.}
                 \label{fig:correl_int_loop_simu}
	\end{figure}



	\begin{table*}
		\caption{Comparison of results from the 1D loop simulation with the observations from \citet{froment2015}}
		\label{table:event_1_time_lag_contour}
		\centering
			\begin{tabular}{c c c c}
				\hline\hline
				\multicolumn2c{Time lag (from cross correlations, in minutes)}   								 &  1D hydrodynamic simulation &  Observations \\

				\hline
				\multicolumn2c{Between $T_e$ and $n_e$  }				  &	111		&      119   \\
				\hline
				\multirow{6}{*}{Between the AIA channels} 			& 335-211  &     109	 	&     113    \\
															& 211-193  & 	 37  	        &     26     \\   
															& 335-193  &	 127		&     145    \\
															& 94-335    &	 -95		&      -115   \\
															& 335-171  &	 122		&      142   \\
															& 171-131  &	 -11		&      -1      \\
				\hline
			\end{tabular}
	\end{table*}	


We perform the same time lag analysis as in \citet{froment2015} with the six light curves of synthetic intensities (from 5~hr after the beginning of the simulation to the end of the simulation), averaged around the loop apex (i.e around $s = 200$~Mm). We cross-correlate six pairs of channels, the cross-correlation values for 94-335, 335-211, 211-195, 335-193, 335-171, and 171-131 can be seen in Figure~\ref{fig:correl_int_loop_simu}. All the peak cross-correlation values are above 0.7, even though the light curves of Figure~\ref{fig:int_loop_simu} can \textit{a priori} look dissimilar (e.g., the 335 and 171 channels). The time lag values are given in Table~\ref{table:event_1_time_lag_contour}. 

According to this method, the peaks in the synthetic intensities follow each other in the order: 335~\AA, 94~\AA, 211~\AA, 131~\AA, 171~\AA, and 193~\AA. This is not exactly the same order as the position of the peaks seen in Figure~\ref{fig:int_loop_simu}. 
The time lags determined by the cross-correlation are influenced not only by the intensity peaks but by the intensity evolution during the whole cycle. These values thus give a general tendency (we witness the cooling of the loop) but do not necessarily give the exact order of the peaks \citep[see also][]{lionello2016}. 

The fact that the EUV loops are generally seen in their cooling phase is well-known \citep[e.g.][]{warren2002, winebarger2003, winebarger_warren2005, ugarte-urra2006, ugarte-urra2009, mulu-moore2011, viall&klimchuk2011}. \citet{viall&klimchuk2012} and \citet{viall&klimchuk2013} argue by means of observations and modeling that the widespread cooling seen in the active regions implies an impulsive character of the heating. However, while our simulation uses a constant heating, we also witness this cooling behavior. As already underlined by \citet{lionello2013} and \citet{lionello2016}, simulations of loops with a quasi-constant heating can reproduce the observed time lags. 

\medskip

If the temperature evolution in time was fully symmetric (e.g. cosine-shaped) and the density were constant, we would expect to see two peaks in each intensity passband (for passbands sensitive to these temperatures), one when the temperature increases and passes the peak response of the channel, and another when the temperature decreases and passes again through the peak response. In this case, we would observe no systematic time delay between the channels. However, if the rise of the temperature were faster than the temperature fall and/or if the density were lower during the heating than during the cooling phase, the first peak would disappear and a systematic correlation would result. The time lags observed between the EUV intensities in the corona are then not necessarily due to the impulsive character of the heating. As seen in Figure~\ref{fig:int_loop_simu}, time lags are observed and the temperature rise is steeper than the fall while variations of the density are delayed compared to the temperature. This explains why the cross correlation picks out on the cooling phase.

	\begin{figure}
		\centering
                 \includegraphics[width=\linewidth]{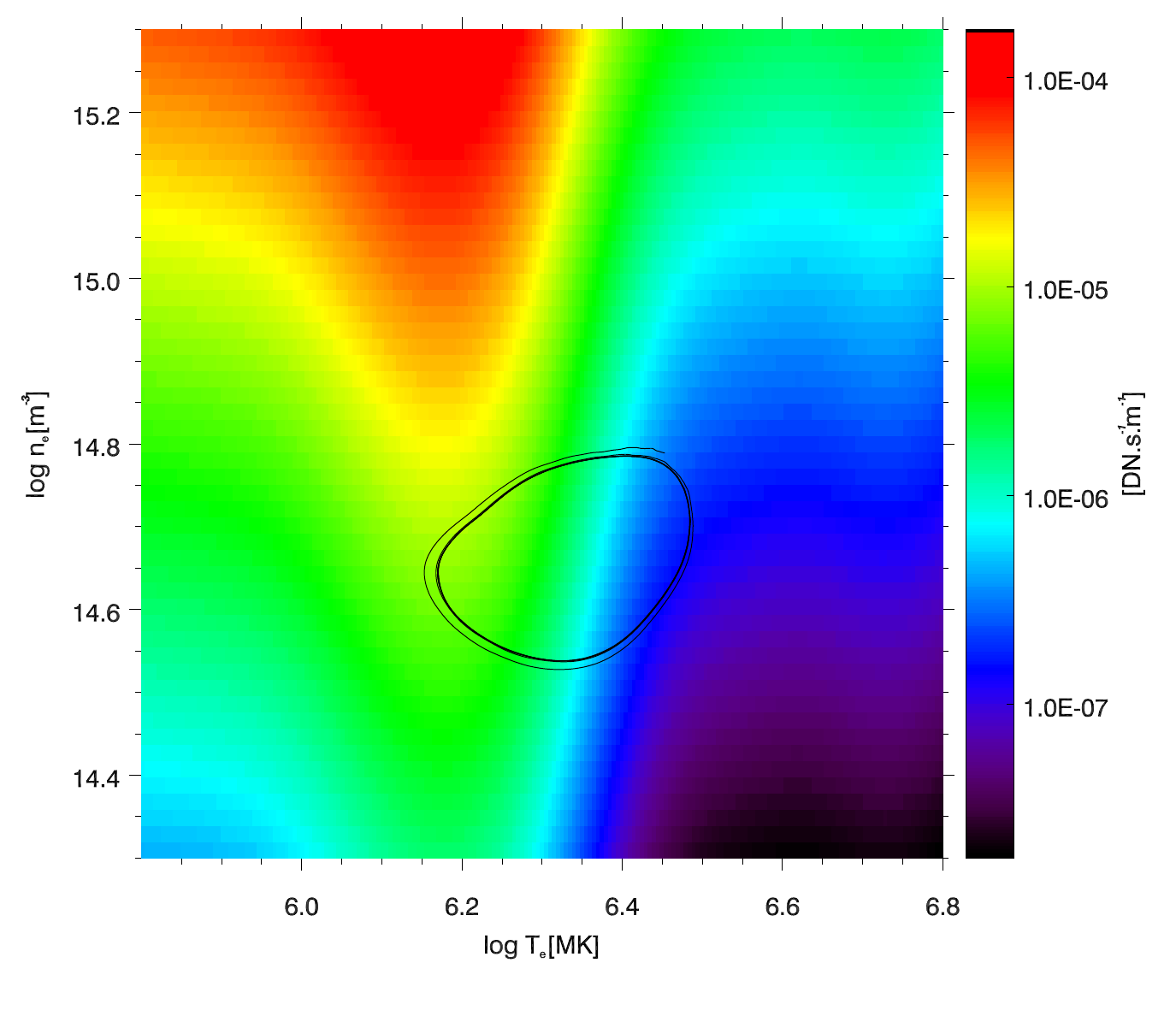}
                 \caption{Phase diagram for the loop simulation. The temperature and density cycle (averaged around the apex) is plotted with in background the map of $G(n_e,T_e)=\frac{1}{4\pi} n_e^2 R_b(n_e,T_e) $, i.e. the volume emissivity of the plasma (without the integration along the line of sight) at 193~\AA. The cycle is counterclockwise.}
                 \label{fig:space_int_loop_simu}
	\end{figure}
	

In Figure~\ref{fig:space_int_loop_simu}, we plot the phase diagram for the loop simulation. We use $T_e$ and $n_e$ averaged around the loop apex, starting 5 hr after the beginning of the simulation. In the background of the cycles, we displayed the map of $G(n_e,T_e)=\frac{1}{4\pi} n_e^2 R_b(n_e,T_e)$, i.e. the volume emissivity of the plasma (without the integration along the line of sight) for the 193~\AA~passband. In Figure~\ref{fig:int_loop_simu} we can see that when the temperature is the highest (above 2.5~MK in absolute value, i.e. $\log T_e = 6.4$), the intensity in the 193~\AA~passband is nearly flat. Indeed, in Figure~\ref{fig:space_int_loop_simu}, the ($T_e,n_e$) cycle spans an area of low $G(n_e,T_e)$ above $T_e=2.5$~MK ($\log T_e = 6.4$). We can see that if the cycle were somewhere else in the ($T_e,n_e$) domain, the intensity cycle response would be quite different. We emphasize that the time lags resulting from intensity variations are sensitive to the combination of $T_e$ and $n_e$ variations and should be interpreted carefully.

	\subsubsection{Comparison with the thermal properties of the observed pulsating loops}

	
	\begin{figure*}  
		\centering
                \includegraphics[width=\linewidth]{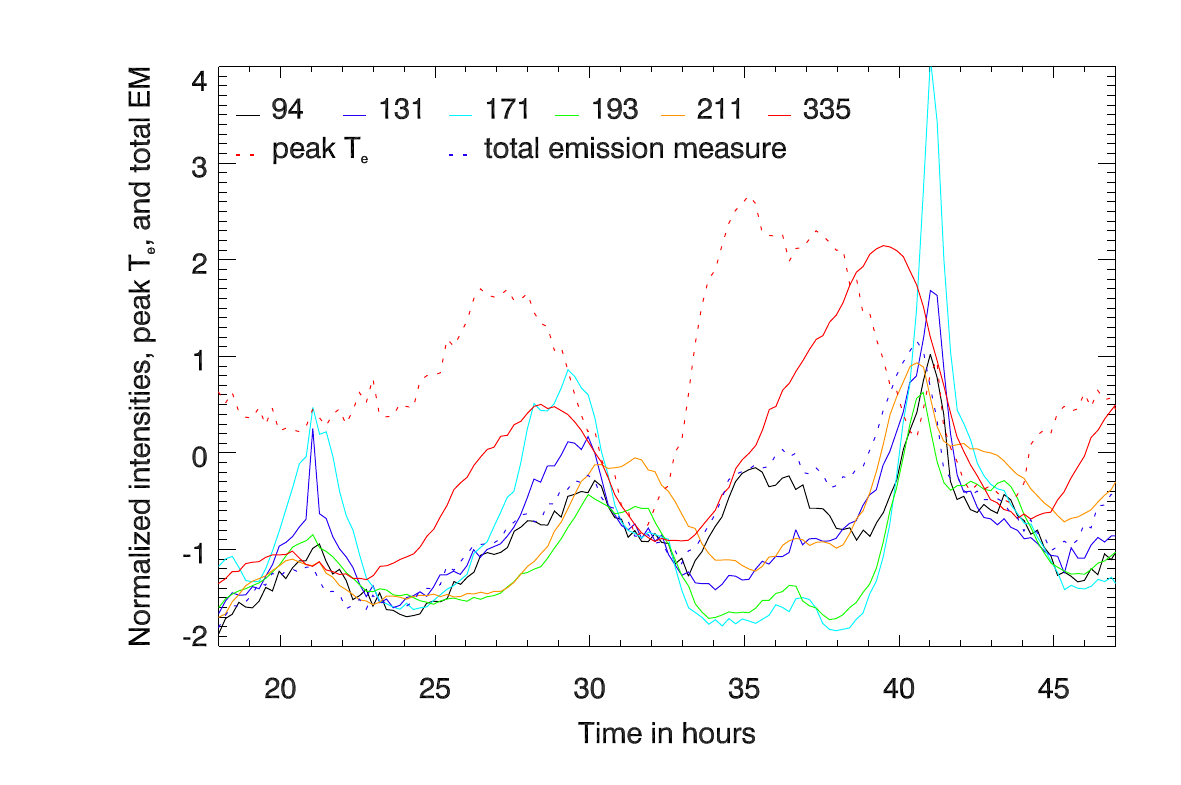}
                 \caption{Observed intensity time series (13 minutes of cadence) in the six coronal channels of AIA \citep[adapted from Figure 2 of ][]{froment2015}. For each channel, the intensity is averaged over a contour close to the loop bundle apex \citep[see Figure 1 of][]{froment2015}. We zoom on two cycles between 18 hr and 47 hr after the beginning of the sequence. In dotted lines, we overplot the evolution of the peak temperature (in red) and of the total emission measure (in blue), averaged over the same contour. The intensities and DEM parameters are all normalized to their standard deviation (we substract the mean curve and divide by their standard deviation).}
                 \label{fig:int_loop_obs}
	\end{figure*}
		
	
We now compare the properties of the synthetic intensity time series with the ones of the observed loops in \citet{froment2015}.

In Figure~\ref{fig:int_loop_obs}, we present the evolution of the intensity in the six channels of AIA from observations of event~1 in \citet{froment2015}. These intensities (at a 13-minute cadence) are averaged over a contour close to the loop bundle apex  \citep[see Figures 1 and 2 of ][]{froment2015}. They are plotted in the same way as in Figure~\ref{fig:int_loop_simu}: we zoom on two cycles between 18 hr and 47 hr after the beginning of the sequence. We add also the evolution of the peak temperature and of the total emission measure from the DEM analysis of \citet[][see Figure 2]{froment2015}.

Comparing Figure~\ref{fig:int_loop_simu} and Figure~\ref{fig:int_loop_obs}, we can see that the long-term behavior (the 9~hr cycles) is very well reproduced by our simulation. However in the simulation, the short-term dynamics is missing. As discussed previously, our simulation has the inherent limitations of 1D simulations: only a single loop is studied, with no background emission. The simulation also assumes a perfectly constant heating, so that it cannot reproduce the small-scale temporal variations.

The behavior of the light curves during the heating phase (i.e. the nearly flat intensity evolution) that is seen in the simulation is also present in the observations. We can also see that, as in the simulation, the shape of the 335~\AA~light curve and ones of the other bands are different (see at a time around 40~hr in the observations of Figure~\ref{fig:int_loop_obs}).

The evolution of the peak temperature of the DEM and of the total emission measure in the observations seem also to be consistent with the behavior seen in our simulation. However, the total emission measure evolution is noisy, probably due to the poor constraint on the DEM slope \citep[see the discussion in][]{froment2015}.

\medskip

In Table~\ref{table:event_1_time_lag_contour}, we present a comparison between the time lags measured for the observed loops \citep{froment2015} and for the simulated one. With a same period of intensity pulsations (9 hr), the time lags between the AIA channels for the observed and the simulated loop are quite similar, given the inherent limitations of this type of analysis, as mentioned above for the differences between the cross-correlation method and the manual determination of the time lags between peaks. Moreover, as highlighted in \citet[see Table~1]{froment2015}, the effects of the background and foreground emission could change the channel ordering. We also need to bear in mind that we are comparing our observational results with the simulated intensities of a single 1D loop. This is likely to be the biggest source of differences between our observations and simulations.

As already noted in Figure~\ref{fig:phy_param_simu} and Figure~\ref{fig:int_loop_simu}, we find by cross-correlations a time delay between the temperature and the density with a value of 111~minutes. This time delay corresponds to a peak cross-correlation value greater than 0.9. The time delay between the measured DEM peak temperature and the total emission measure for the observed pulsating loops \citep{froment2015} is similar: 119 minutes.

Therefore the thermal properties of the simulated loop, as the intensities evolution,  are consistent with those of the observed pulsating loops of \citet{froment2015}.

\section{Relationship with coronal rain}\label{sec:discussion}

Coronal rain was first observed by \citet{kawaguchi1970} and \citet{leroy1972}. \citet{antiochos1991} proposed a model for the formation of prominences with a heating concentrated at low altitudes. In \citet{antiochos1999}, the authors argue for a common formation mechanism for prominences and coronal rain, namely the loss of thermal equilibrium, giving an explanation for the formation of mass condensations in the corona.

However, while TNE is the common physical process to explain coronal rain, the periodic appearence of coronal rain showers that involves long periods (several hours) in simulations, has never been observed yet. Since coronal rain is often observed for durations of less than several hours, it is impossible to detect a potential periodicity with about the same period.
On the other hand, while several studies suggest that coronal rain and, by extension, TNE in the corona is a phenomenon that could be more common than generally thought \citep[e.g.,][]{antolin_observing_2012,antolin2012,antolin2015}, a proper quantification of coronal rain is still lacking.

The common pulsating behavior of EUV coronal loops, as reported by \citet{auchere2014}, brings a new interesting element to the understanding of coronal rain and of the heating of coronal loops. The detailed study of the thermodynamic behavior of three typical events detected in loops shows that these observations are evidence for TNE in warm coronal loops \citep{froment2015, auchere2016b}. These long-period EUV intensity pulsations, just like coronal rain, are due to a TNE state and are thus an observational manifestation of coronal heating mechanisms. Therefore this widespread behavior observed in loops tells us both about the localisation of the heating in loops and about the heating time scale. 

\medskip

Coronal rain appears to be a different manifestation of the same phenomenon studied here, except that the condensations, locally in the corona, seem to be stronger in the case of coronal rain. 
The long-period intensity pulsations observed with the coronal channels of AIA and studied in the present paper can indeed be modeled using a incomplete condensation scenario. This model can explain the average behavior of the loop bundle observed. That does not exclude the presence of coronal rain in these loops, i.e. localized complete condensations. AIA cannot constrain the behavior of the plasma at transition region and chromospheric temperatures \footnote{However, it is worthnoting that we do not we use the 304 channel because we do not detect pulsations in this passband.}. Moreover, the spatial resolution is not sufficient with AIA to observe coronal rain easily \citep{antolin2015}, especially on the disk.

The temperature of these condensations could for example drop to chromospheric temperatures, with the rest of the plasma remaining at coronal temperatures. It is also possible that we observed only the beginning of a progressive cooling. In fact, temperatures at a part of the western leg of the simulated loop do drop to transition region temperatures (see Figure~\ref{fig:phy_param_simu}). This model is not inconsistent with the observations of \citet{straus_steady-state_2015}, i.e. downflows to only transition region temperatures and not to chromospheric ones. Further work is needed to understand the nature of the condensations associated with the long-period intensity pulsation events detected.

\medskip
	
An alternative interpretation to TNE to explain the observations of long-period intensity pulsating loops might be the self-organization of loops to a state of marginal collisionality, as simulated by \citet{imada2012}. In this paper, the authors simulated coronal loops ($\sim$~50~Mm long) with a 1D hydrodynamic code, testing the model proposed by \citet{uzdensky2007} and \citet{cassak2008}. They assumed that the loops are heated by magnetic reconnections with a density-dependent heating rate. The plasma is self-regulated, with footpoint evaporations and downflow evacuations. This behavior causes pulsations in the temperature and density of the loops with periods of severals tens of minutes.
These periods are smaller than the ones that we observed but the simulated loops are shorter too. However, even if the plasma behavior has several similarities with the one we observed in \citet{froment2015}, the temperature and density evolutions are anti-correlated, which is not what we observe.
An analysis of the velocity of the flows in the observations could be another way to distingush between the two models.

\bigskip

\section{Summary}\label{sec:summary}

In \citet{froment2015}, the physical behavior of long-period EUV intensity loop events was examined. We concluded that these show evidence for cycles of evaporation and incomplete condensation and thus for TNE. 

In this paper, we use an 1D hydrodynamic simulation with a realistic loop geometry from LFFF extrapolations.
We choose an highly-stratified and asymmetric heating function to reproduce the period of pulsation and the temperature deduced from the observations. 
With this 1D hydrodynamic description we are able to reproduce the long-term variations of our observations for NOAA AR 11499 \citep{froment2015}, i.e., the intensity pulsations in the six coronal channels of AIA (with 9 hr of period) with realistic time lags between the channels around the loop apex. 
This supports an explanation of long-period intensity pulsations in terms of TNE.

\citealp{viall&klimchuk2011, viall&klimchuk2012,viall&klimchuk2013} argue that the time lags, and thus the cooling observed in active regions, are consistent with nanoflare storm models involving low-frequency nanoflares. During such nanoflare storms many nanoflare events occur along the structure considered. The low-frequency here implies that the time delay between two heating events is longer than the typical cooling time. This allows time for the plasma to cool before it is reheated.
In our case, we use a constant heating, mainly localized at low altitude, but we witness the same cooling behavior. This was also highlighted in 3D simulations with a quasi-constant and highly stratified heating in \citet{lionello2013,lionello2016} and \citet{winebarger_investigation_2016}. 
It is thus worth noting that high-frequency heating models are also consistent with the time lags observed.
The observed cooling alone is thus not sufficient to discriminate between the heating mechanisms.

For the present loop geometry, we can only reproduce the cycles of evaporation and condensation with a highly-stratified and asymmetric heating profile. Moreover, we found a low-lying null point and many bald patches around the eastern loop footpoint. This topology, which is not seen around the western footpoint, represents a potential site of preferential reconnection at one footpoint, leading to asymmetric heating, which is a key ingredient to reproduce our observations. The sensitivity of the production of such cycles to both loop geometry and heating geometry will be examined in a future paper. 

In conclusions, our simulation further strengthens the interpretation of the observed pulsations in terms of TNE, as proposed by \citet{froment2015}.
 
\acknowledgements
This work is an outgrowth of the work presented during the VII Coronal Loop Workshop and at Hinode 9. The authors acknowledge useful comments from attendees of these conferences. The authors would like to thank Jim Klimchuk for fruitful discussions on thermal nonequilibrium and long-period pulsations in loops, and the referee for constructive comments that contributed to improve the paper. The SDO/AIA and SDO/HMI data are available by courtesy of NASA/SDO and the AIA and HMI science teams. This work used data provided by the MEDOC data and operations centre (CNES / CNRS / Univ. Paris-Sud), \url{http://medoc.ias.u-psud.fr/}. Zoran Miki\'c was supported by NASA Heliophysics Supporting Research Grant NNX16AH03G.
     
\bibliographystyle{aasjournal}                       
\bibliography{biblio}

\end{document}